\documentclass[12pt]{article}
\usepackage{amssymb}
\usepackage{graphicx}
\usepackage{amsmath, amsthm}
\usepackage{hyperref}
\usepackage{subfigure}

\usepackage{fullpage}
\usepackage[font=small]{caption}
\usepackage{setspace}
\newcommand{\asss}[2]{\noindent{\textbf{Assumption #1.} #2}}

\newtheorem{theorem}{Theorem}
\newtheorem{prop} [theorem] {Proposition}

\input{tcilatex}
\begin{document}

\title{Multivariate Specification Tests Based on a Dynamic Rosenblatt Transform
}
\author{\textsc{Igor L. Kheifets} (ITAM)\\ \\
\url{https://kheifets.github.io/}
\\ \\{\small Accepted to Computational Statistics and Data Analysis}
\thanks{
{\textcopyright} 2018. This
manuscript version is made available under the CC-BY-NC-ND 4.0 license
\url{http://creativecommons.org/licenses/by-nc-nd/4.0/}
}
}
\date{\today}
\maketitle


\begin{abstract} \footnotesize This paper considers parametric model adequacy
tests for nonlinear multivariate dynamic models. It is shown that commonly used
Kolmogorov-type tests do not take into account cross-sectional nor
time-dependence structure, and a test, based on multi-parameter empirical
processes, is proposed that overcomes these problems.  The tests are applied to
a nonlinear LSTAR-type model of joint movements of UK output growth and
interest rate spreads.  A simulation experiment illustrates the properties of
the tests in finite samples.  Asymptotic properties of the test statistics
under the null of correct specification and under the local alternative, and
justification of a parametric bootstrap to obtain critical values, are
provided. 

\textbf{Keywords}: Diagnostic test, joint distribution, multivariate modeling,
Rosenblatt transform, LSTAR model.

\textbf{JEL classification}: C12, C22, C52.
\maketitle

\end{abstract}

\section{Introduction}

Robust nonparametric methods are hard to implement in a multidimensional case,
and parametric modeling is often called for. For example, linear and nonlinear
VAR models with Gaussian innovations are often used in macroeconometrics, while
multivariate volatilities, which can be described by different types of
multivariate GARCH (MGARCH) or copula-based models, are popular in financial
econometrics. The use of a misspecified parametric model may result in
misleading conclusions, in particular, biased estimates of monetary policy
effects and underestimation of the risk in financial models.  Thus it is
crucial to develop specification testing procedures for these models. In a
multidimensional context, it is important to know not only the time structure
of the random vectors but also the dependence between contemporaneous
variables, and this dependence should be used, for example, for portfolio
diversification. Hence, we should be able to test for the correct specification
of the joint multivariate distribution conditional on past information.

There is a huge literature on testing multivariate normality, see Mecklin and
Mundform (2004). For testing a general type of distribution in a dynamic setup,
a dynamic version of the Rosenblatt Transform (cf. Rosenblatt 1952), which is
also a type of a Probability Integral Transform, PIT, allows us to approach all
kinds of distributions in a unified manner. The idea is that, given the true
conditional distribution, one can transform the data to independent and
identically distributed (i.i.d.) uniform random variables and possibly further
to normal i.i.d.  Then, instead of testing the shape of the initial
distribution, the uniformness and independence of the transformed data can be
evaluated with histograms and correlograms, as suggested by Diebold et al.
(1999) and Clements and  Smith (2002) for multivariate density forecast
evaluation.  In practice, however, the distribution is known only up to
parameters; therefore the method of Diebold et al. (1999) can not be applied.
The reason is that, when estimates are plugged in, the (dynamic) Rosenblatt
Transform delivers only approximately i.i.d.\ uniforms;  moreover, the
asymptotic distribution may change, and even become model and case dependent,
see Durbin (1973). Ignoring parameter uncertainty introduces severe size
distortions in such tests, as documented in simulations by Kalliovirta and
Saikkonen (2010). The problem usually is solved by either transforming the
statistics of interest to make them convergent to a known  distribution
(Khmaladze 1981, Wooldridge 1990, Delgado and Stute 2008, and many others) or
by approximating critical values by the parametric bootstrap (Andrews 1997).
For nonparametric testing of multivariate GARCH models, Bai and Chen (2008)
developed a Kolmogorov-type testing procedure based on the dynamic Rosenblatt
Transform. They applied a Khmaladze (1991) martingale transformation
(K-transformation) to the empirical process to obtain a limiting distribution
of statistics. However, in a univariate setup, Corradi and Swanson (2006) noted
that Kolmogorov-type tests, based on a one-parameter empirical process, may not
distinguish some important alternatives to the conditional distribution. In an
i.i.d. setup with conditioning on covariates, Delgado and Stute (2008) proposed
a consistent test using a two-parameter empirical process coupled with a
Khmaladze martingale transformation. In a time series setup, where a
Kolmogorov-type test does not capture misspecification in the dynamics,
Kheifets (2015) proposed a test based on a multi-parameter empirical process
and used a bootstrap to obtain critical values.

In this paper, we consider nonparametric testing of a multivariate distribution
specification. We study the consequences of using a Kolmogorov-type test in
this setup. We find that Kolmogorov-type tests result in missing both dynamic
and cross-section dependence. To overcome this problem, we consider tests based
on a multivariate empirical process, adapting the weak convergence results of
Kheifets (2015) to a multivariate case. As well as a Kolmogorov test, our
PIT-based procedure can test nonlinear models and capture deviations in
marginal distribution. Besides that, our test includes two ingredients: a
dynamic check (similar to Kheifets 2015) and a cross-section check.  Thus our
technique may be used not only for testing but also for investigating sources
of misspecification.  Our test complements the parametric tests of Kalliovirta
and Saikkonen (2010) and Gonzalez-Rivera and Yoldas (2012).  We avoid bandwidth
selection, and our test statistics have a parametric rate of convergence,
unlike the smoothing techniques of Hong and Li (2005), Li and Tkacz (2011), and
Chen and Hong (2014), who use kernels to estimate the conditional distribution
function and spectrum. 

The contribution of the paper is the following: We develop the test and apply
it to a model of UK growth and interest spreads and perform a set of Monte
Carlo experiments to study the performance of the test in finite samples, where
we observe results similar to Kolmogorov tests' in some cases and improvement
in others.  We derive asymptotic properties of the test under the null and the
local alternative, taking into account the parameter estimation effect, and
justify the use of bootstrapped critical values. 

The rest of the paper is organized as follows. Section~2 introduces
specification test statistics, based on the dynamic Rosenblatt Transform. The
empirical application is in Section~3.  Monte Carlo experiments are shown in
Section~4.  Section~5 concludes.  Asymptotic properties of the test are listed
in the Appendix. 

\section{The Test Statistics}

\subsection{Our proposal}

We now explain our methodology in detail. Suppose that a sequence of $d\times
1$ vectors $Y_{1},Y_{2},\ldots,Y_{T}$, where $Y_{t}=(Y_{t1},Y_{t2},
\ldots,Y_{td})^{\prime }, t=1,...,T,$ is given. Let $\Omega _{t}$ be the
information set at time $t$ (not including $Y_{t}$), i.e., the $\sigma$-field
of $\{Y_{t-1},Y_{t-2},\ldots \}$.

We consider a family of joint distributions $F_{t}(y|\Omega _{t},\theta )$,
conditional on the past information, parameterized by $\theta \in \Theta $,
where $\Theta \subseteq R^{L}$ is a finite dimensional parameter space.  Apart
from allowing the conditional (information) set $\Omega _{t}$ to change with
time, we permit change in the functional form of the distribution using
subscript $t$ in $F_{t}$. Our null hypothesis of correct specification is:
\bigskip

$H_{0}$ : The multivariate distribution of $Y_{t}$ conditional on $\Omega
_{t} $ is in the parametric family $F_{t}(y|\Omega _{t},\theta )$ for some $
\theta _{0}\in \Theta $.

\bigskip

Note that by specifying the multivariate conditional distribution, we specify
many properties of the data simultaneously, such as multivariate and
univariate marginal distributions, univariate conditional distributions, time
and cross-section dependence, symmetry, all existing moments, kurtosis, etc.
Many dynamic models can be written in the form of a conditional distribution.
Examples include (non)linear vector autoregressive (VAR) and MGARCH models with
i.i.d. parametric innovations, copula-based models with parametric marginals and
possibly time-varying copula functions, and discretely sampled continuous-time
models represented by a stochastic differential equation.

We now describe how to use PIT. In a simple univariate unconditional
testing, we have that if $Y\sim F(\cdot)$, then $U=F(Y)$ is uniform, which is
the base of the Kolmogorov test. More precisely, the null hypothesis of the
Kolmogorov test is that $F(Y)$ is uniform. If we are interested in
conditional distribution testing, which is the case when we have covariates or
dynamics, we use the fact that if $Y_t|\Omega_t\sim F_t(\cdot|\Omega_t)$, then
$U_t=F(Y_t|\Omega_t)$ is uniform and i.i.d. The distribution needs to be
absolutely continuous. For a discrete distribution, one may use a different
transform, see Kheifets and Velasco (2013, 2017). 

In multivariate setup, $F(Y),$ which seems to be an obvious generalization of
the PIT, is not generally uniform for a vector $Y\sim F(\cdot )$. $F(Y)$ is
related to copula functions, which describe multivariate dependence without
specifying marginals. The properties of $F(Y)$ were studied in Genest and
Rivest (2001). In this paper we want to check the specification of the
multivariate distribution; therefore, testing the specification of the copula
function is not sufficient. For a multivariate PIT, we need to define 
the conditioning sets properly. Following Rosenblatt (1952) and Diebold et al.
(1999), construct a long univariate series by stacking sequentially $Y_{t}$ to
get a
$n=Td$ long univariate series, which we denote
\begin{equation}\label{eq:defz}
\{Z_{k}\}_{k=1}^{n}:=\{...,Y_{t1},Y_{t2},...,Y_{td},...\}_{t=1}^{T}.
\end{equation}
In other words, $Z_{k}=Y_{t\ell}$ for $t=\lceil k/d\rceil$ and $\ell=k-td$,
where $k=1,...,n$, $t=1,...,T$, $\ell=1,...,d$, and $\lceil x\rceil$ denotes
the smallest integer not less than $x$.  There are many ways to order and stack
$Y_{t\ell}$; for example,
$\{...,Y_{t2},Y_{t1},...,Y_{td},...\}$ is another possibility. Although the
ordering determines the test statistics, in practice we are not able to test
all possible orderings.
 In our Monte Carlo experiments,
we study how the choice of a particular order affects the performance of the
test.  From the null multivariate conditional distributions $F_{t}(y|\Omega
_{t},\theta _{0})$, we can obtain univariate conditional distributions $
F_{Z_{k}}(z|\Omega _{k}^{Z},\theta _{0})$, where $\Omega _{k}^{Z}$ is a
$\sigma$-field of $\{Z_{1},Z_{2},...,Z_{k-1}\}$. To do this, for each $t$,
apply to the joint distribution $d$ times the factorization $F\left(
y_{1},y_{2}\right) =F\left( y_{1}\right) F\left( y_{2}|y_{1}\right) $. Now
apply $n$ times PIT
\begin{equation*}
U_{k}=F_{Z_{k}}(Z_{k}|\Omega _{k}^{Z},\theta _{0}),
\end{equation*}
which are uniform and i.i.d. This is a dynamic analog of a multivariate
Rosenblatt (1952) transform. Explicit formulas of such transforms for VAR
models with possible MGARCH innovations can be found in Bai and Chen (2008) and
Kalliovirta and Saikkonen (2010); for copula-based models see Patton (2013),
see also \ref{sect:transform} for an example.

Under the null, $\{U_{k}\}_{k=1}^{n}$ are uniform and i.i.d. Diebold et al.
(1999) use this fact for density forecast evaluation by looking at
histograms and correlograms of $U_{t}$. In this paper, we use the ideas of
Delgado and Stute (2008) and Kheifets (2015) to make a formal testing procedure
for $H_0$ that requires us to check
simultaneously uniformness and independence. For  $k=2,...,n$, using
pairwise independence and uniformness, we have
\begin{equation*}
\Pr ({U}_{k}\leq r_{1},{U}_{k-1}\leq r_{2})=r_{1}r_{2},
\end{equation*}
for $r\in {[0,1]}^{2},$ which motivates us to consider the following
empirical processes
\begin{equation}
{V}_{2T}(r)=\frac{1}{\sqrt{n-1}}\sum_{k=2}^{n}\left[ I({U}_{k}\leq r_{1})I({U
}_{k-1}\leq r_{2})-r_{1}r_{2}\right] .
\end{equation}
If we do not know $\theta _{0}$, we approximate $U_{k}$ with
\begin{equation*}
\hat{U}_{k}=F_{Z_{k}}(Z_{k}|\Omega^Z_{k},{\hat{\theta}}),
\end{equation*}
where ${\hat{\theta}}$ is an estimator of $\theta _{0}$, so that
\begin{equation}
\hat{V}_{2T}(r)=\frac{1}{\sqrt{n-1}}\sum_{k=2}^{n}\left[ I(\hat{U}_{k}\leq
r_{1})I(\hat{U}_{k-1}\leq r_{2})-r_{1}r_{2}\right] .  \label{eq:Vpn}
\end{equation}
To obtain a test statistic, define
\begin{equation*}
D_{2T}=\Gamma (\hat{V}_{2T}(r)),
\end{equation*}
for any continuous functional $\Gamma (\cdot )$ from $\ell ^{\infty
}([0,1]^{2})$, the set of uniformly bounded real functions on $[0,1]^{2}$,
to $R$. In particular, we consider Cramer-von Mises (CvM) and
Kolmogorov-Smirnov (KS) statistics
\begin{equation}\label{eq:D2}
D_{2T}^{CvM}=\int_{[0,1]^{2}}\hat{V}_{2T}(r)^{2}dr\text{ and }
D_{2T}^{KS}=\sup_{[0,1]^{2}}\left\vert \hat{V}_{2T}(r)\right\vert .
\end{equation}
The choice of the measure is an interesting question in itself; it may make the
test more powerful for some particular alternatives. Here we stick just to
these two.

To check $p$-wise independence in a similar way to Delgado (1996), we can
base test statistics on the $p$-parameter process
\begin{equation*}
{V}_{pT}(r)=\frac{1}{\sqrt{n-p}}\sum_{k=p}^{n}\left[ \prod_{j=1}^{p}I({U}
_{k-j+1}\leq r_{j})-r_{1}r_{2}\ldots r_{p}\right],
\end{equation*}
using norms on $[0,1]^p,$ say,
\begin{equation}\label{eq:Dp}
D_{pT}^{CvM}=\int_{[0,1]^{p}}\hat{V}_{pT}(r)^{2}dr\text{ and }
D_{pT}^{KS}=\sup_{[0,1]^{p}}\left\vert \hat{V}_{pT}(r)\right\vert .
\end{equation}
The
test based on the process with $p=1$ does not have power against many
important alternatives, e.g., omitted autoregressive terms in mean or variance.
Such alternatives are captured by tests with $p=2$.  In the rest of the paper
we consider such tests. 

To test other than one-lag pairwise dependence, for $j=1,2,...$ define the
process
\begin{equation*}
\hat{V}_{2T,j}(r)=\frac{1}{\sqrt{n-j}}\sum_{k=j+1}^{n}\left[ I(\hat{U}
_{k}\leq r_{1})I(\hat{U}_{k\text{-}j}\leq r_{2})-r_{1}r_{2}\right] ,
\end{equation*}
with test statistics,
\begin{equation}\label{eq:D2j}
D_{2T,j}^{CvM}=\int_{[0,1]^{2}}\hat{V}_{2T,j}(r)^{2}dr\text{ and }%
D_{2T,j}^{KS}=\sup_{[0,1]^{2}}\left\vert \hat{V}_{2T,j}(r)\right\vert .
\end{equation}
In order to provide some guidance for a practitioner, we propose the following
interpretation of the test statistics above.
For example, for linear bivariate model $D_{2T,1}$ controls cross-sectional
dependence, while $D_{2T,2}$ controls dynamic specification. In DGP-A1, DGP-C5
and DGP-C6 in the Monte Carlo section below, cross-sectional dependence is
misspecified. Such alternatives are captured by $D_{2T,1}$. In DGP-A3, the
dynamics in the first variable is misspecified, hence $D_{2T,2}$ catches it.
Marginal checks are included in both statistics, i.e.,~misspecification of the
marginal distribution will result in high values for all test statistics. To
test the null hypothesis against an unknown alternative, aggregate information
from these statistics. Following the notation of Kheifets (2015), define for
$k=1,...,n-1$ the test statistics
\begin{equation*}
ADJ_{kT}=\sum_{j=1}^{k}D_{2T,j}^{CvM}\text{ and }MDJ_{kT}=
\max_{j=1,...,k}D_{2T,j}^{KS},
\end{equation*}
and
\begin{equation*}
ADJ_{kT}^{0}=D_{1T}^{CvM}+ADJ_{kT}\text{ and }MDJ_{kT}^{0}=\max \left(
D_{1T}^{KS},MDJ_{kT}\right) .
\end{equation*}
Test statistics $ADJ_{kT}$ and $MDJ_{kT}$ summarize the information from
different lags, in the spirit of a Ljung-Box (1978) test, for which choosing
small $k$ may miss higher order dependence, while choosing large $k$ reduces
overall power. Test statistics $ ADJ_{kT}^{0}$ and $MDJ_{kT}^{0}$ additionally
take into account the marginal distribution of the data explicitly.

Since, under $H_{0}$, we know (up to the parameter value) the parametric
conditional distribution, we apply a parametric bootstrap to mimic the $H_{0}
$ distribution based on $F_{t}(\cdot |\cdot,\hat{\theta})$. This will require
additional computational time; however, note that even in the case of asymptotic
distribution free tests, bootstrap critical values may be preferred to
asymptotic ones in terms of finite sample performance, see Kilian and Demiroglu
(2000). We introduce the algorithm now.

\begin{enumerate}
\item Estimate the model with the original data $Y_{t}$, $t=1,2,...,T$ to get
parameter estimator $\hat{\theta}$ and test statistic $\Gamma (\hat{V})$.

\item Simulate $Y_{t}^{\ast }$ with $F_{t}(\cdot |\Omega _{t}^{\ast },\hat{
\theta})$ recursively for $t=1,2,...,T$, where $\Omega _{t}^{\ast
}=(Y_{t-1}^{\ast },Y_{t-2}^{\ast },...)$.\label{basimulate}

\item Estimate the model with simulated data $Y_{t}^{\ast }$ to get $
\theta ^{\ast }$ and bootstrapped statistics $\Gamma (\hat{V}^{\ast })$.

\item Repeat 2-3 $B$ times, and compute the percentiles of the empirical
distribution of the $B$ bootstrapped statistics.

\item Reject $H_{0}$ if $\Gamma (\hat{V})$ is greater than the $(1-\alpha )$
th percentile of the empirical distribution.
\end{enumerate}

We recommend setting $B$ to at least $1000$, the number used in this paper.
Bootstrapping critical values takes 15 minutes for the bivariate LSTAR model
considered in Section~\ref{sect:emp}, see~\ref{sect:comp} for further
computational details.   For some bootstrap samples the estimation routine may
fail.  For example, when an explicit solution to a model exists, numerical
methods may not be able to invert an ill-conditioned matrix.  In case of highly
nonlinear models, extremum estimation procedures may take a long time to
converge or even to return an error. In these cases, $B$ should be larger so that
the bias in the bootstrapped quantiles, which may appear after omitting those
samples, is acceptable to a researcher.  

Theoretical properties of the procedure developed in this paper rely on 1) the
dynamic extension of the Rosenblatt transform and 2) the weak convergence
result in Kheifets (2015).  The latter proposes a specification test for
univariate models based on multi-parameter empirical processes and studies the
convergence of such processes, which allows us to justify the use of the
bootstrapped critical values. These theoretical results can be employed for our
needs because the Rosenblatt transforms are univariate (and uniformly and
independently distributed under the null hypothesis). In order to do so, we
 need to verify high-level assumptions of Kheifets (2015) under the dynamic
extension of the Rosenblatt transform, please see \ref{sect:verifya} for details.

\subsection{Comparison with Bai and Chen (2008) and related tests}

Most of the tests for $H_0$ use some particular properties of the null
distribution, such as skewness, kurtosis or correlation, whether of the initial
data or of that of transformed variables (i.e., residuals, PITs or composition
of PITs, and transforms to normal random variables). One exception is the test
of Bai and Chen (2008), which is  a nonsmooth omnibus-type test for
multivariate GARCH models. In fact, they can test our $H_0$.  Their test
statistic is based on a combination of empirical processes
\begin{equation*}
{J}_{T,\ell}(r)=\frac{1}{\sqrt{T}}\sum_{t=1}^{T}\left[ I({U}_{(t-1)d+\ell}\leq
r)-r\right] , \end{equation*}
for $r\in[0,1]$, $\ell=1, \ldots,d$.  In particular, they propose three
  combinations: taking  the maximum, $\max_{\ell=1,
  \ldots,d}\sup_{r\in[0,1]}|{J}_{T,\ell}(r)|$, taking the sum, $\sum_{\ell=1,
  \ldots,d}\sup_{r_{\ell}\in[0,1]}|{J}_{T,\ell}(r_{\ell})|$, and pooling
  $\sup_{r\in[0,1]}|\frac{1}{\sqrt{d}}\sum_{\ell=1,\ldots,d} {J}_{T,\ell}(r)|$,
  corresponding to $D_{1T}^{KS}$ in our notation.  They further apply a
  K-transformation to obtain the convergence to the supremum of a standard
  Brownian motion, even in the presence of parameter estimation.  However, these
  types of processes miss the dynamic check in $U_k$, i.e., the critique raised
  by Corradi and Swanson (2006) for univariate tests applies in a multivariate
  case as well. The key element of our procedure is to use the test statistics
  based on the multivariate processes $D_{pT}$, $p>1$.  In the next section, we
  compare the performance of the test statistics $D_{1T}$ and $D_{2T}$ using
  Monte Carlo simulations.

Another common strategy is to bootstrap CvM or KS statistics based 
on the following ``row-wise" empirical process (e.g., Patton~2013, Section~4.1, 
but note a typo in the standardization,
should be $\frac{1}{\sqrt{T}}$ instead of $\frac{1}{{T}}$)
\begin{equation*}
{S}_{pT}(r)=\frac{1}{\sqrt{T}}\sum_{t=1}^{T}\left[ \prod_{\ell=1}^{d}I({U}
_{(t-1)d+\ell}\leq r_{j})-r_{1}r_{2}\ldots r_{d}\right],
\end{equation*}
for $r\in[0,1]^d$. Unlike the tests based on $J_{T,\ell}$, 
tests based on this empirical process do take some dynamics in $U_k$ into
account, but not all. In particular, independence is controlled \emph{within}, 
but not \emph{between}, the  rows $U^t=({U}_{(t-1)d+1},{U}
_{(t-1)d+2},\ldots,{U}
_{(t-1)d+d})$. For instance, in a bivariate case, (cross-sectional) 
independence between $U_{2t-1}$ and $U_{2t}$, $t=1,\ldots,T$ is controlled, 
while (dynamic-cross) independence between $U_{2t}$ and $U_{2t+1}$
and (pure dynamic) independence between $U_{2t-1}$ and $U_{2t+1}$ and 
between $U_{2t}$ and $U_{2t+2}$ are not. 
The empirical process $V_{pT}$, suggested in our paper, overcomes these
problems. 

We calculate critical values using a parametric bootstrap. In a similar
(although univariate) situation, Corradi and Swanson (2006) suggest the use of
a block bootstrap, which delivers correct critical values under dynamic
misspecification. Our approach is different in two respects.  First, we do not
require strict stationarity, while their block bootstrap approximation relies
on it.
Second, in our case dynamic misspecification is not allowed under the null and
must be detected if present. For example, predictions of output growth and
spreads considered in our empirical application are based on all available
information. That is why the conditional set in our null hypothesis is
$\Omega_t$ and not $Y_{t-1}$.  

Bai and Chen (2008) apply K-transformation to obtain a distribution-free test.
Extending these kinds of transformations to our case is non-trivial because the
empirical processes are multivariate and the same variables enter different
dimensions, introducing dependence among dimensions. A parametric bootstrap is
an attractive alternative to such techniques because it requires repeated
simulation and estimation of the null model, while implementing a K-transform
requires one to derive analytically the transform for each model under
consideration, which is not straightforward (see examples in Bai and Chen,
2008).    

\section{Empirical Application}\label{sect:emp}

\begin{figure}[!t]
\centering 
  \includegraphics[width=0.50\textwidth]{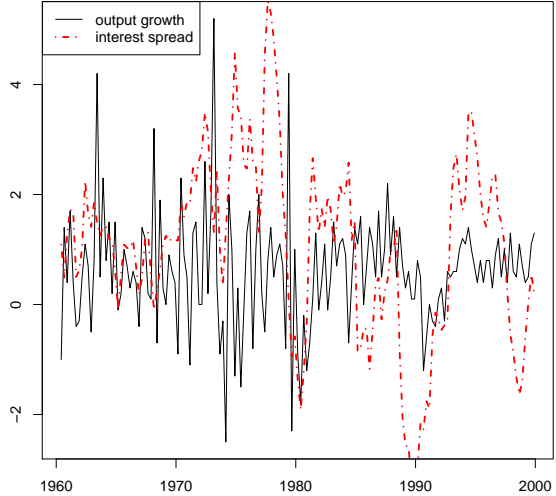}
\caption{UK output growth and interest spread, quarterly data from 1960:3 to
1999:4} \label{fig:data}
\end{figure}

We test a model for joint movement of output growth and the interest rate
spread in the UK, plotted on Figure~\ref{fig:data} and suggested by Anderson,
Athanasopoulos, and Vahid (2007, AAV hereafter). Output growth ($Y_{t1}$) is
calculated as $100$ $\times$ the difference of logarithms of seasonally
adjusted real DGP, and the spread ($Y_{t2}$)  is the difference between the
interest rates on 10 Year Government Bonds and 3 Month Treasury Bills. The
sample consists of $159$ quarterly time series observations, dating from 1960:3
to 1999:4 and available, together with the GAUSS programs, at the AAV
accompanying webpage
(\url{http://qed.econ.queensu.ca/jae/2007-v22.1/anderson-athanasopoulos-vahid/}). 

AAV suggested the following model for the conditional mean of $Y_t$,
$\mu_t=\left(\mu_{t1},\mu_{t2}\right)'$  as a function of unknown parameters
$b=\left(b_1,\ldots,b_{11}\right)'$
\begin{eqnarray*}
\mu_{t1} &=&b_1+b_2 Y_{t-2,1} + b_3 Y_{t-3,2} + b_4 Y_{t-1,2} 
+w_t \left(-b_5 Y_{t-1,2}\right), \\
\mu_{t2} &=& b_8+b_9 Y_{t-3,1} + b_{10} Y_{t-1,2}+ b_{11} Y_{t-2,2}, \\
\text{where } w_{t} 
&=& \left(1+\exp\left(-b_6(Y_{t-2,1} - b_7)\right)\right)^{-1}.
\end{eqnarray*}
The logistic smooth transition autoregressive specification for output growth
incorporates different regimes and smooth transitions between them.

AAV estimated the model by ML and obtained predicted values by simulation,
using (conditional) normality of errors. This motivates us to test the
following null hypothesis
\begin{equation}\label{eq:H0NAR}
H_{0}:Y_{t}|\Omega_t\sim N(\mu_{t} ,\Sigma ), \text{ where
$\Sigma$ is independent of $\Omega_t$.}
\end{equation}
Assuming~(\ref{eq:H0NAR}), we obtain $\hat
b=\left(0.35,0.21,0.15,0.32,-0.52,2.56,0.68,0.20,-0.14,1.14,-0.26\right)'$ and
$\Sigma=\left(0.95,-0.02;-0.02,0.45\right)$. AAV report only the estimate of
$b$,  not $\Sigma$. Our estimate of $b$ is close to theirs. Tests' $p$-values
are reported in Table~\ref{tbl:emp}. As in AAV, we detect no remaining serial
correlation. However, the nonlinear procedure proposed in our paper suggests
that the  null hypothesis (\ref{eq:H0NAR}) can be rejected at the $1\%$
confidence level.

\begin{table}[t]
\caption{\footnotesize$100$ $\times$ $p$-values of tests for the null hypothesis
defined in Section~\ref{sect:emp} applied to the UK data. Both Cramer-von Mises
and Kolmogorov-Smirnov norms are used.
Tests based on the one-parameter process,
the two-parameter process with 1st lag for cross sectional dependence, and
the two-parameter process with 2d lag for time dependence,
as well as Ljung-Box (Box et al., 1994) test statistics with $1$, $2$, $3$,
$20$, and $25$ lags, are considered. Sample size is $T=156$. }
\centering
\setlength{\tabcolsep}{1.8pt}\footnotesize
\begin{tabular}{ccccccccccccc}
\hline
 &  & $ D_{1}^{CvM} $  & $ D_{2,1}^{CvM} $  & $ D_{2,2}^{CvM} $  & $ D_{1}^{KS} $  & $ D_{2,1}^{KS} $  & $D_{2,2}^{KS} $  & $ LBQ_1 $  & $ LBQ_2 $  & $ LBQ_3 $  & $ LBQ_{20} $  & $ LBQ_{25} $  \tabularnewline
\hline\hline
&   & $ 0.2 $  & $ 0.5 $  & $ 0.0 $  & $ 0.0 $  & $ 0.0 $  & $ 0.0 $  & $ 31.3 $  & $ 54.9 $  & $ 68.1 $  & $ 31.4 $  & $ 48.3 $  \tabularnewline
  \hline
 \end{tabular}
 \label{tbl:emp}
\end{table}

Monte Carlo experiments conducted in the next section show that the empirical
size of our tests is close to nominal. Simulations C2, C3, and C4 in the next
section also suggest that alternative models with Student-$t$ distribution and
the same dynamics as in~(\ref{eq:H0NAR}) can hardly be captured by serial
correlation checks, while our tests have power for sample size $T=100$.

\section{Monte Carlo Experiments}\label{sect:mc} 

We study the performance of our procedure under the null and alternative
hypotheses using Monte Carlo experiments discussed in Giacomini, Politis and
White (2013). We also show the rejection rates for parametric tests Ljung and
Box (1978) and Jarque and Bera (1980) (for Part C).  For comparison purposes,
their distributions are obtained using bootstrap.  We start with a simple
linear null hypothesis and linear DGP, where model estimation is simple and
estimators converge fast, allowing us to see the finite sample behavior of the
proposed statistics;  then we proceed to a nonlinear model motivated by our
empirical example and calibrated to real data.  All DGPs are listed in
Table~\ref{tbl:dgps}.  The number of Monte Carlo repetitions is set to $1000$.

\begin{table}[p]\label{tbl:dgps}
\caption{The list of DGP used in Monte Carlo simulations}
\begin{enumerate}
\item DGP-A1: $Y_{t}|\Omega_t\sim N\left( \left(0,0\right)' ,\left(2 , \alpha;\alpha, 1\right) \right)$
\item DGP-A2: $Y_{t}|\Omega_t\sim t_5\left( \left(0,0\right)' ,\left(2 , \alpha;\alpha, 1\right) \right)$
\item DGP-A3: time series with lag-$1$ dependence are normally distributed
$Y_{t}|\Omega_t\sim N\left(A_{1}Y_{t-1},(1 , 0;0 , 1)\right)$, where $A_{1}=(\alpha , 0; 0 , 0)$.
\item DGP-B1: $Y_{t}|\Omega_t\sim N\left( \left(0,0\right)' ,\left(1 , 0.5;0.5 , 1\right) \right) $,
\item DGP-B2: $Y_{t}|\Omega_t\sim N\left( \left(0,0\right)' ,\left(1 , 0.5;0.5 , 2\right) \right) $,
\item DGP-B3: $Y_{t}|\Omega_t\sim N\left( \left(0,0\right)' ,\left(2 , 0.5;0.5 , 1\right) \right) $,
\item DGP-B4: $Y_{t}|\Omega_t\sim t_5\left( \left(0,0\right)' ,\left(1 , 0.5;0.5 , 1\right) \right) $,
\item DGP-B5: $Y_{t}|\Omega_t\sim t_5\left( \left(0,0\right)' ,\left(1 , 0.5;0.5 , 2\right) \right) $,
\item DGP-B6: $Y_{t}|\Omega_t\sim t_5\left( \left(0,0\right)' ,\left(2 , 0.5;0.5 , 1\right) \right) $,
\item
DGP-C1: $Y_t$ is generated by the model considered in Section~\ref{sect:emp},
with $b$ and $\Sigma$  equal to the estimate obtained therein.  
\item
DGP-C2: $Y_t$ as in DGP-C1, but with conditional distribution Student $t_7$ instead of normal.
\item DGP-C3: $Y_t$ as in DGP-C1, but with conditional distribution Student $t_5$ instead of normal.
\item DGP-C4: $Y_t$ as in DGP-C1, but with conditional distribution Student $t_3$ instead of normal.
\item DGP-C5: $Y_t$ as in DGP-C1, but with conditional mean for the output $\tilde\mu_{t1}=\mu_{t1}+ 0.5 Y_{t-2,2}$.
\item DGP-C6: $Y_t$ as in DGP-C1, but with conditional mean for the output $\tilde\mu_{t1}=\mu_{t1}+ 0.9 Y_{t-2,2}$.
\end{enumerate}
\end{table}

\subsection{Part A}
In this subsection, we  study the  finite sample performance of the test
statistics to test the null hypothesis that the series is identically
multivariate normally distributed and is time
and cross-section independent,
\begin{equation}\label{eq:H0A}
H_{0}:Y_{t}|\Omega_t\sim N(\mu ,\Sigma ), \text{ where
$\mu ,\Sigma$ are independent of $\Omega_t,$ and $\Sigma$ is diagonal}.
\end{equation}
We study this null hypothesis under three types of alternatives:
\begin{enumerate}
  \item DGP-A1: series are cross-section dependent, normally distributed,
  \item DGP-A2: series are cross-section dependent, $t_5$ distributed, and
  \item DGP-A3: series are time dependent, normally distributed,
\end{enumerate}
where in all cases the parameter $\alpha$ takes values $0,0.1,...,0.9.$
The value $\alpha=0$ generates data under the null in DGP-A1 and DGP-A3, so
empirical power should be around nominal. Increasing $\alpha$ makes deviation
from the null stronger, so the power should increase. 

\begin{figure}[!t]
\centering 
\subfigure[CvM against DGP-A1] {
  \includegraphics[width=0.30\textwidth]{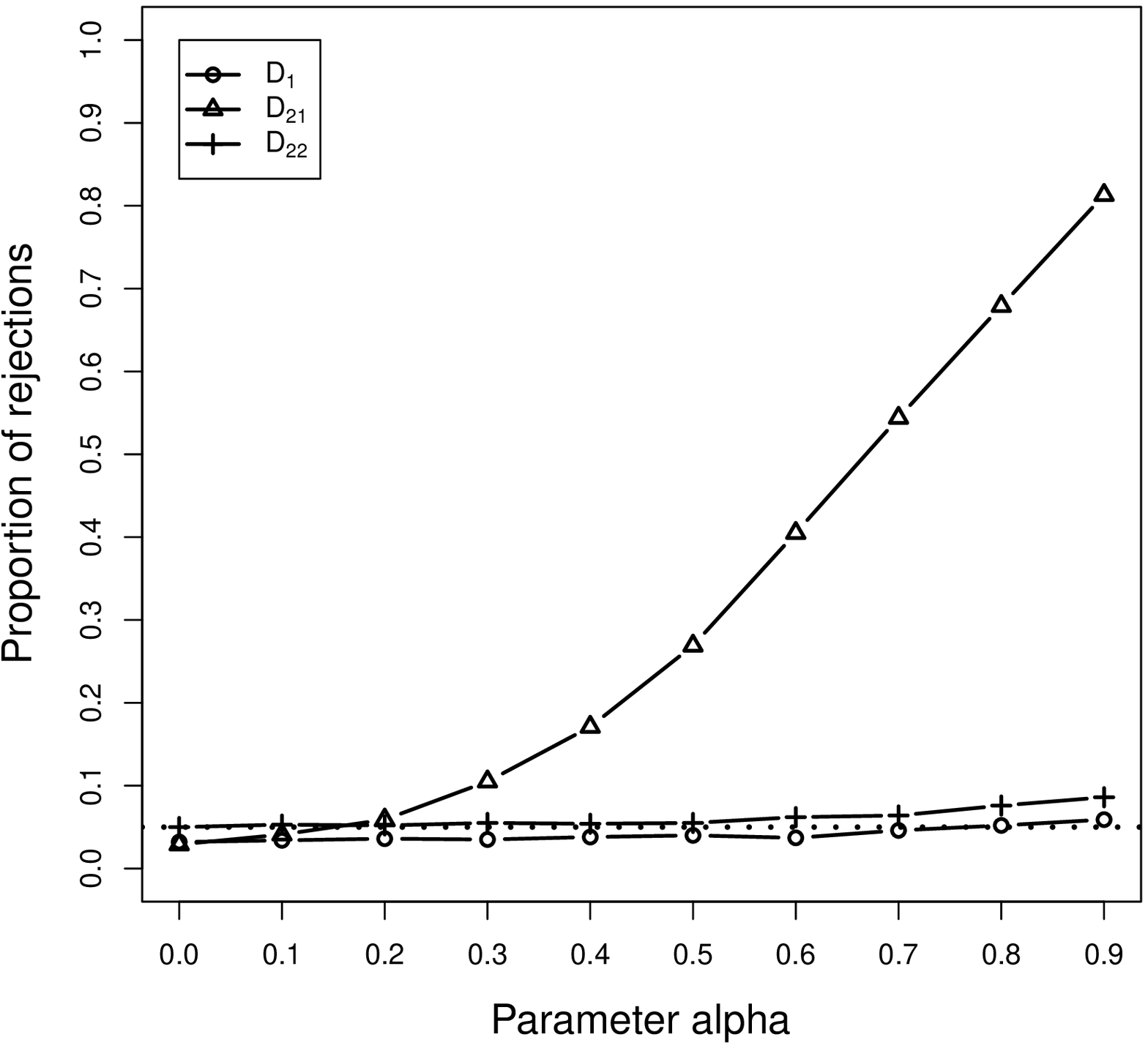}
}
\subfigure[KS against DGP-A1] {
  \includegraphics[width=0.30\textwidth]{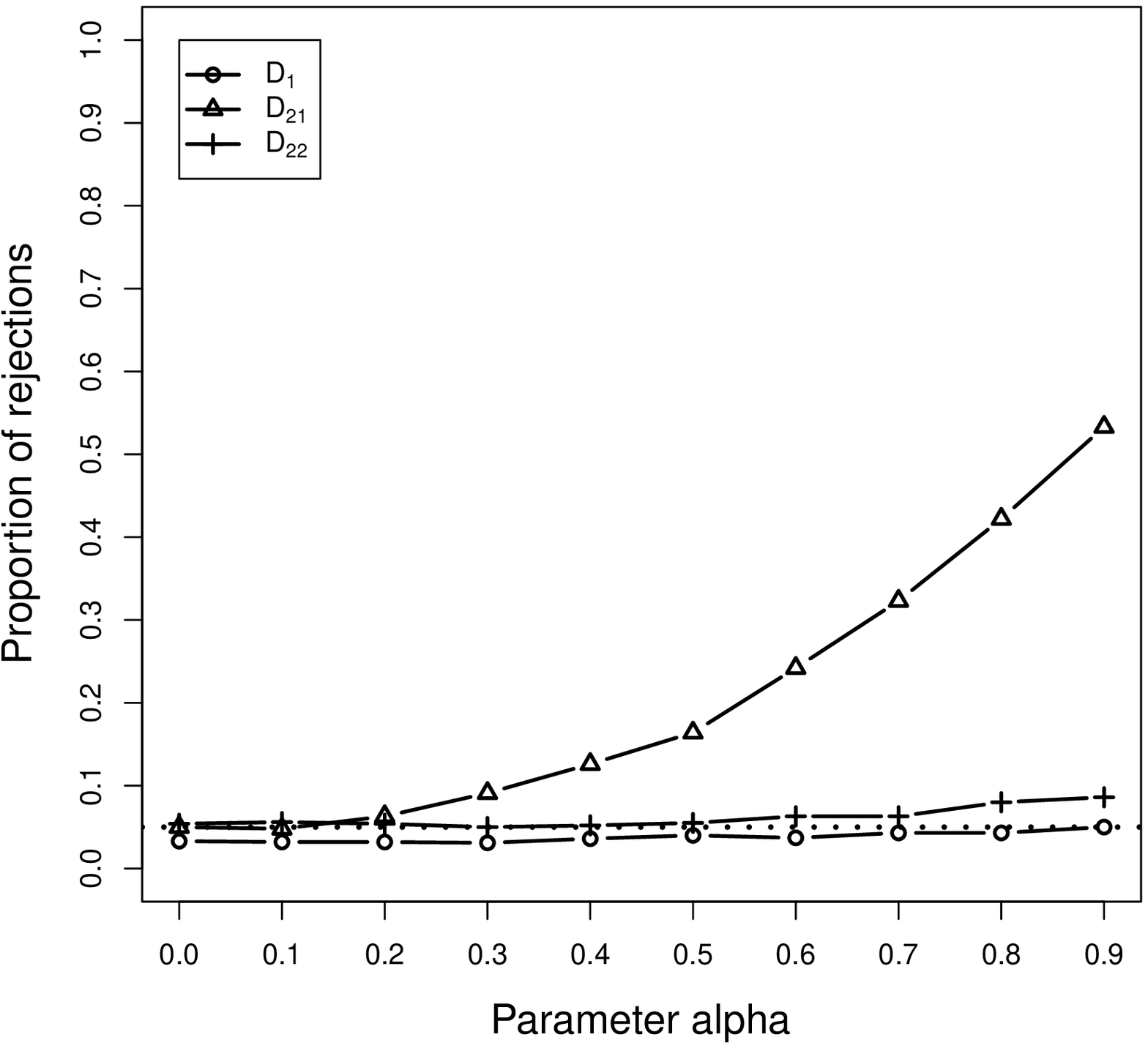}
}
\subfigure[CvM against DGP-A2] {
  \includegraphics[width=0.30\textwidth]{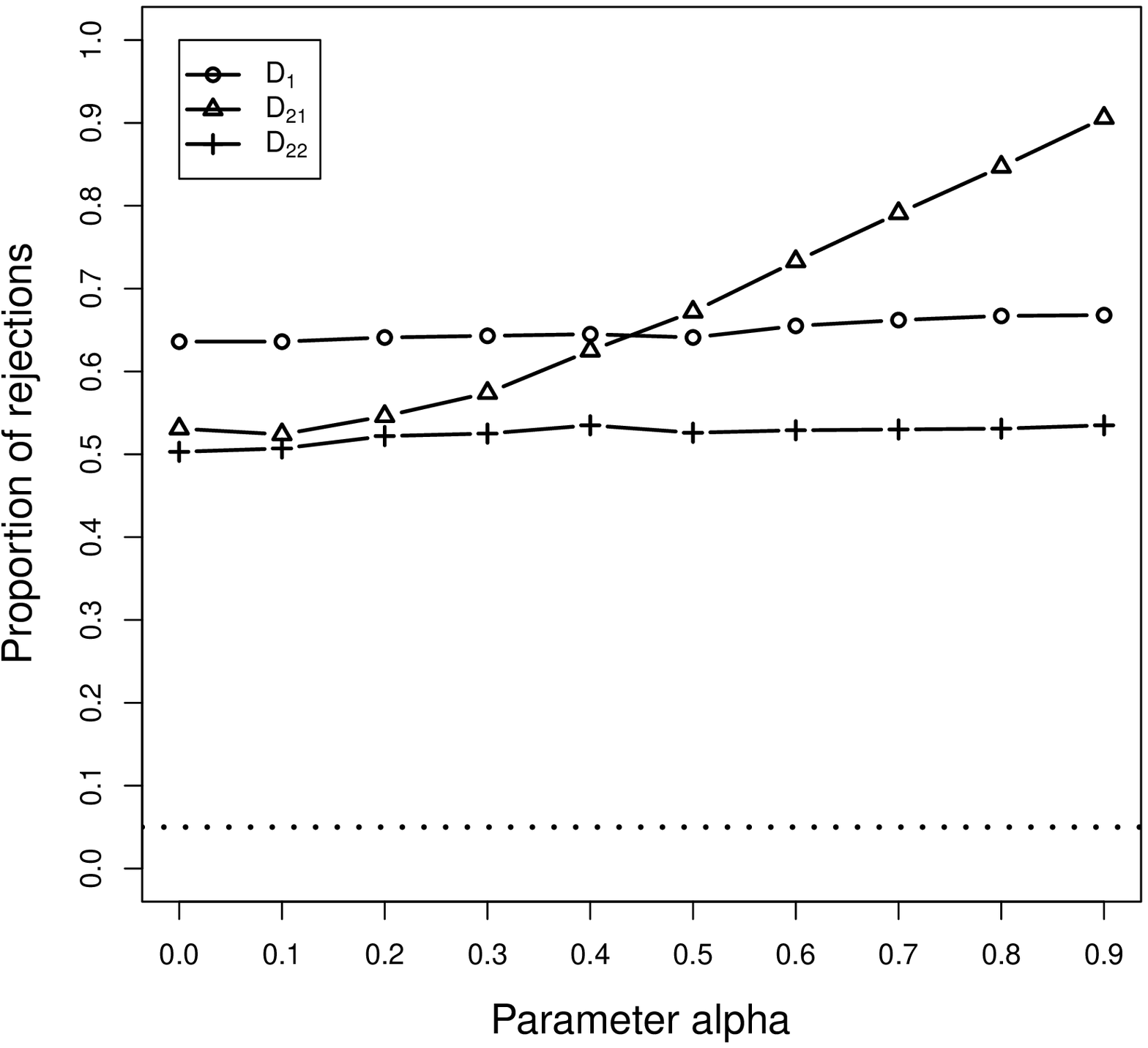}
}
\subfigure[KS against DGP-A2] {
  \includegraphics[width=0.30\textwidth]{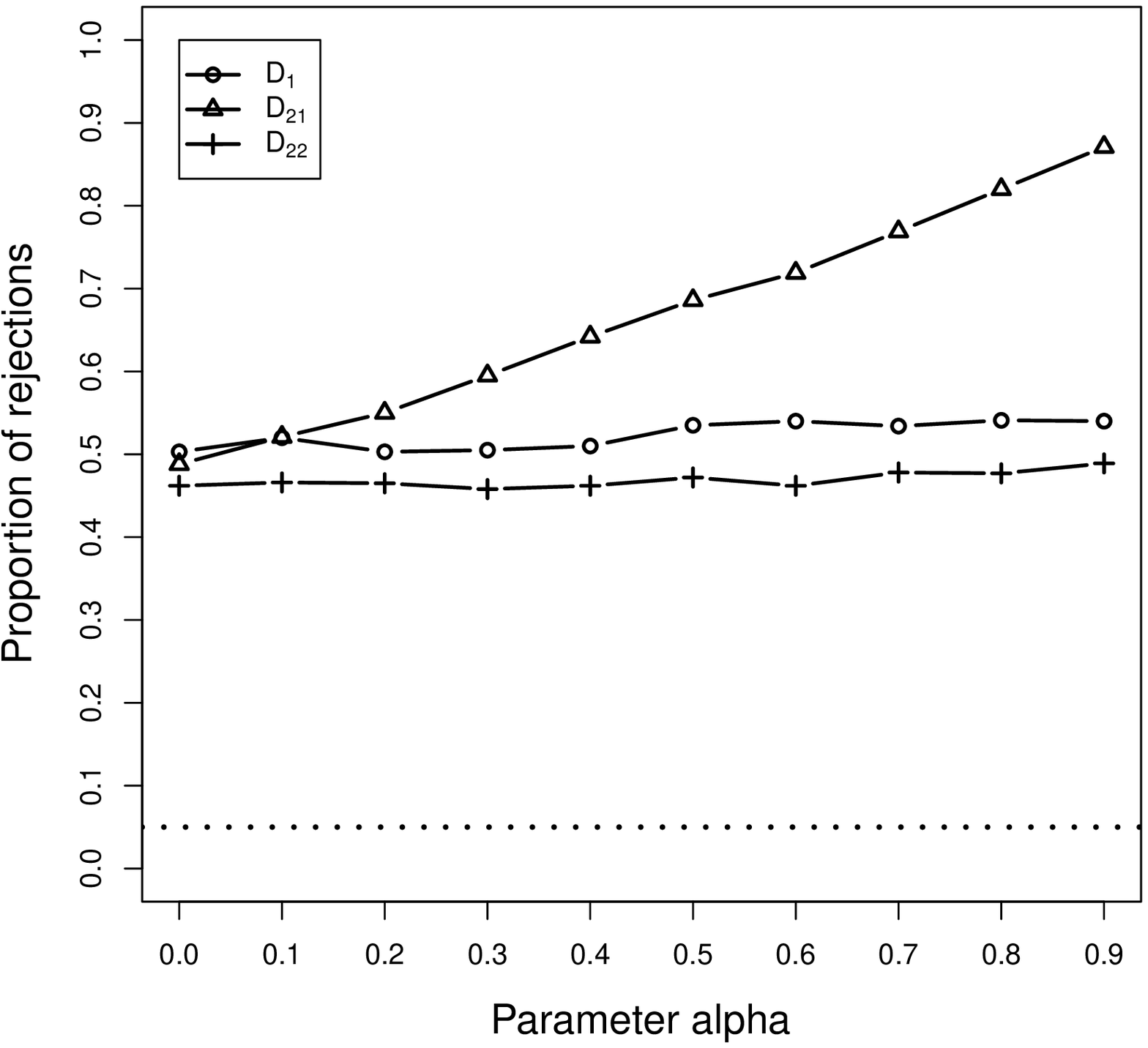}
} 
\subfigure[CvM against DGP-A3] {
  \includegraphics[width=0.30\textwidth]{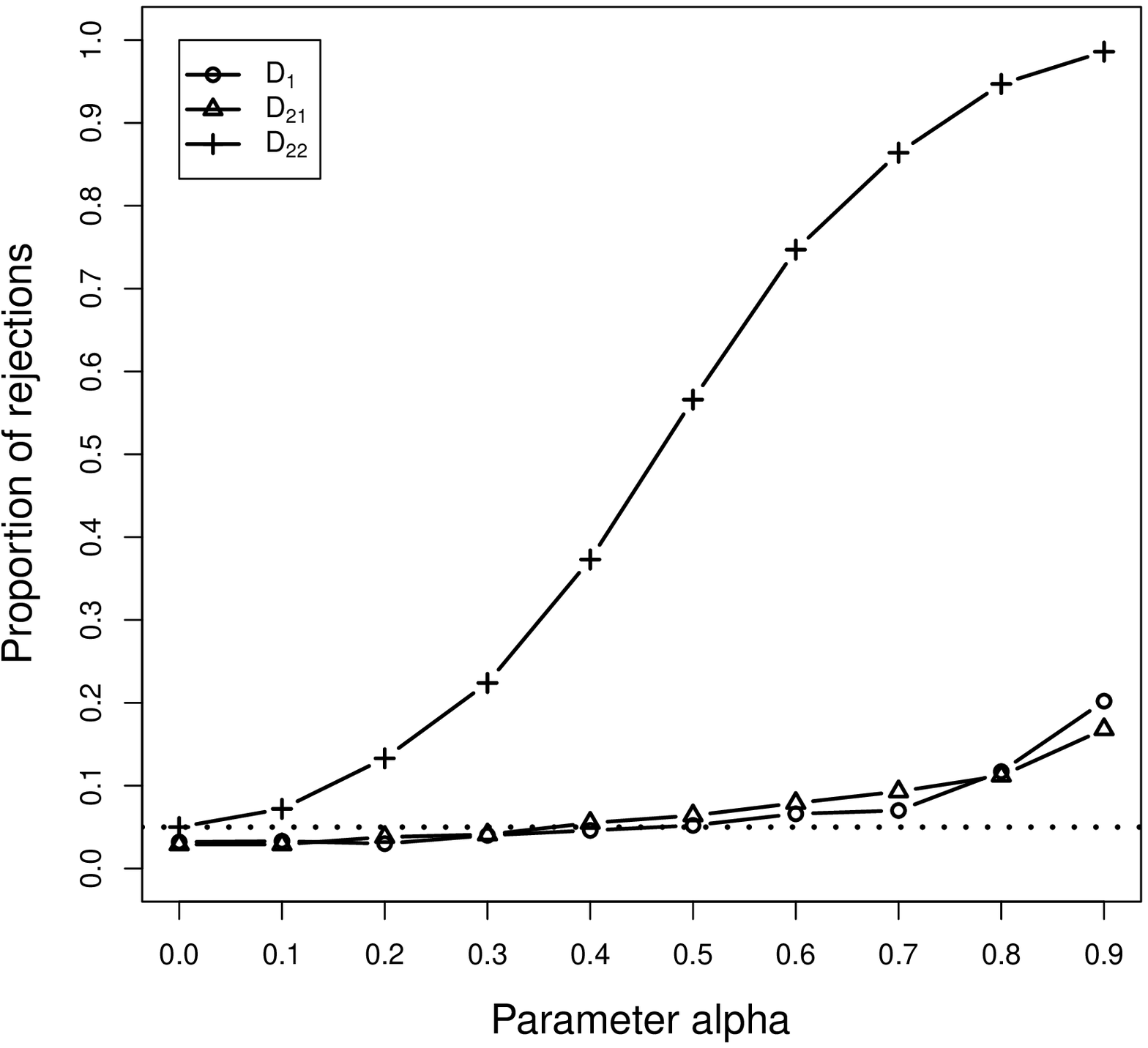}
}
\subfigure[KS against DGP-A3] {
  \includegraphics[width=0.30\textwidth]{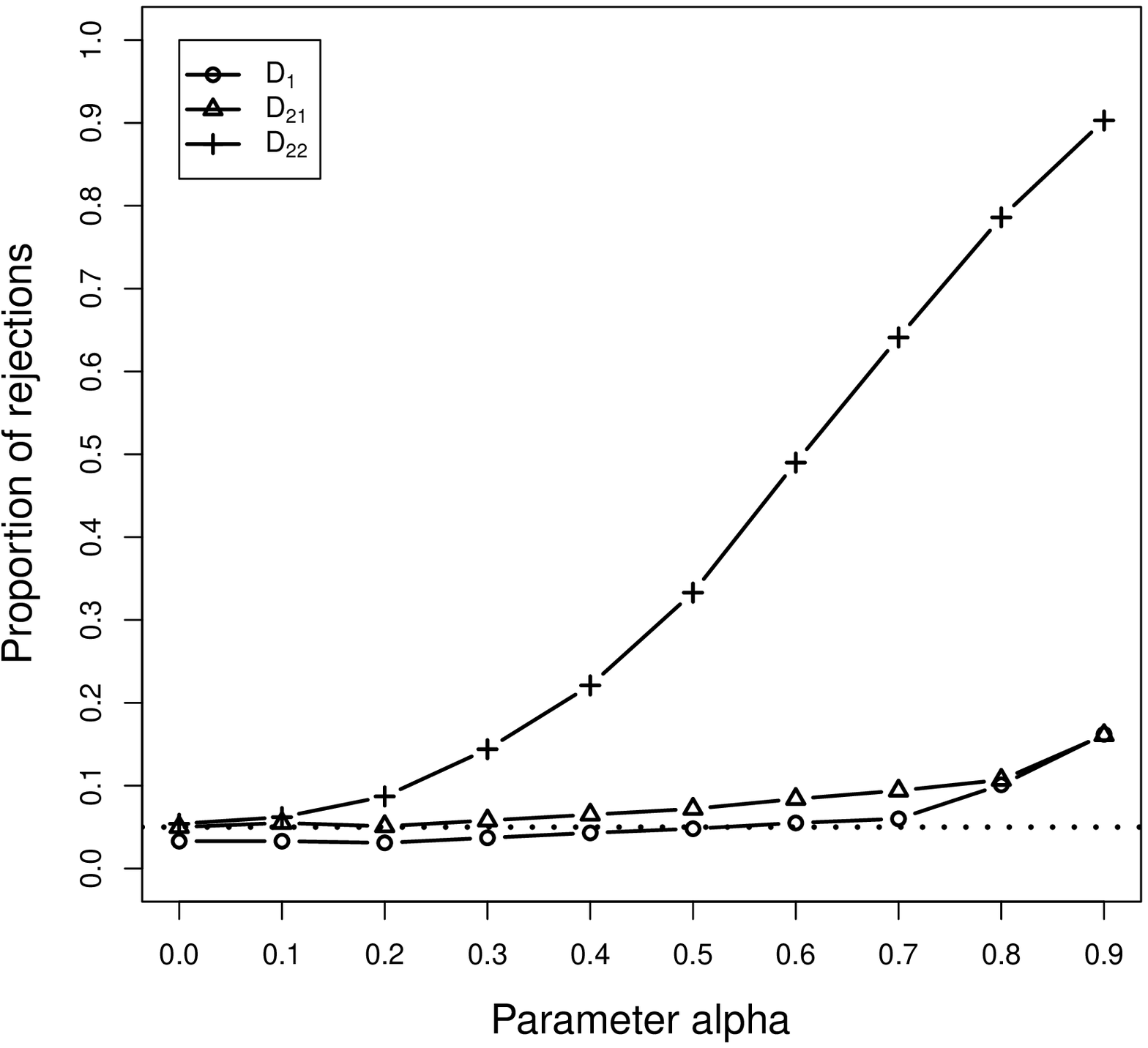}
}
\caption{Rejection proportions of tests for Normal time and cross-section
independent null hypothesis (\ref{eq:H0A}). Panel (a) shows the performance of
the
Cramer-von Mises tests for cross-section dependent normal DGP-A1. Panel (b)
shows the performance of the Kolmogorov-Smirnov test for the same GDP. Panels
(c) and (d) show the performance of the corresponding tests for Student-t with
5 degrees of freedom $t_5$ DGP-A2.  Panels (e) and (f) show the performance of
the corresponding tests for cross-section independent normal VAR(1) DGP-A3.
Tests based on one-parameter process (circle markers),
two-parameter process with the first lag for cross sectional dependence
(triangular markers) and
two-parameter process with the second lag for time dependence
(cross markers) are considered. Rejections at $5\%$ are plotted with a dotted
line. Sample sizes are $T=100$. }
\label{fig:mc1}
\end{figure}

We study empirical power of the tests $D_1$, $D_{2,1}$ and $D_{2,2}$, defined
in (\ref{eq:Dp}) and (\ref{eq:D2j}).  In Figure \ref{fig:mc1}(a), $5\%$
rejection rates of Cramer von Misses tests for DGP-A1 are plotted.  We see that
both $D_1$ and $D_{2,2}$ have near nominal $5\%$ power for all $\alpha$.  Test
$D_{2,1}$, on the contrary, is able to detect alternatives starting from small
power at $\alpha=0.2$ and increasing up to almost $80\%$ for $\alpha=0.9$.  In
Figure \ref{fig:mc1}(c), Cramer von Misses tests for DGP-A2 are plotted. All
tests have at least $50\%$ rejection rates for all $\alpha$.  In Figure
\ref{fig:mc1}(e), Cramer von Misses tests for DGP-A3 are plotted.  We see that
both $D_1$ and $D_{2,1}$ have near  $5\%$ power for all $\alpha$.  Test
$D_{2,2}$, on the contrary, is able to detect alternatives starting from small
power at $\alpha=0.2$ and increasing up to more than $90\%$ for $\alpha=0.9$.
The picture does not change if we use a KS-norm.

In sum, we have observed that $D_{2,1}$ captures cross-sectional
misspecification and $D_{2,2}$ is able to capture misspecification in dynamics,
while $D_{1}$ detects neither.  All three tests perform equally well when the
  marginal distribution is misspecified.

\subsection{Part B}

We study the finite sample performance of the test statistics to test the null
hypothesis that the series is multivariate normal,
\begin{equation}\label{eq:H0B}
H_{0}:Y_{t}|\Omega_t\sim N(\mu ,\Sigma ),\quad \text{where} \ \mu
\ \text{and}\ \Sigma\ \text{are independent of}\ \Omega_t.  \end{equation}
Note that now $\Sigma$ need not be diagonal, so we allow cross-sectional
dependence but not temporal.
DGP-B1 to DGP-B3 generate data under the parametric family null hypothesis
(\ref{eq:H0B}) and deliver the empirical size of the test. DGP-B4 to DGP-B6
simulate data under the $t_{5}$ distribution, a popular alternative to
normality that
accounts for fat tails, and deliver the empirical power of the test. DGP-B1 and
DGP-B4 were considered also in Bai and Chen (2008). DGP-B2, DGP-B3, DGP-B5, and
DGP-B6 are taken to assess the influence of the order choice
in~(\ref{eq:defz}).  We
consider the cases both when the distribution is symmetric with respect to the
coordinates and when it is not. When the distribution is not symmetric,
exchanging the coordinates is equivalent to changing the order
in~(\ref{eq:defz}), and we want to know if, and how, it affects the performance
of the test.  In the normal case, this symmetry is reflected by the restriction
on the covariance matrix $\Sigma _{t11}=\Sigma _{t22}$, case DGP-B1,4. 

We study the performance of tests $D_1$, $D_{2,1}$, and $D_{2,2}$, as defined
in (\ref{eq:Dp}) and (\ref{eq:D2j}) and Ljung-Box (Box et al., 1994) test
statistics with $1$, $2$, $3$, $20$, and $25$ lags.  In Table~\ref{tbl:T100},
sample sizes are $T=100$.  DGP-B1 to DGP-B3 are contained in the null
hypothesis and deliver empirical size, which is slightly under nominal size
almost for all process-based tests.  If we increase sample size up to  $T=300$,
see Table \ref{tbl:T300}, the situation improves, and we even get a light
oversize for DGP-B3.  In general, empirical size looks satisfactory. In DGP-B4
and DGP-B5, the data is generated under $t_5$, and this is captured very well
for both $T=100$ and $T=300$.  DGP-B3 is symmetric to DGP-B2, and DGP-B6 is
  symmetric to DGP-B5, and here we can not see a difference in the performance
  by alternating the order in~(\ref{eq:defz}). There is not much difference in
  using KS or CvM norms.

\begin{table}[t]
\caption{Rejection rates of tests for the Normal independent in time 
composite null hypothesis defined in (\ref{eq:H0B})
and DGP-B1 - DGP-B6. Both Cramer-von Mises
and Kolmogorov-Smirnov norms are used.
Tests based on the one-parameter process, the 
two-parameter process with $1$ and $2$ lags,
as well as Ljung-Box (Box et al., 1994) test statistics with $1$, $2$, $3$,
$20$, and $25$ lags, are considered. Sample sizes are $T=100$. }
\centering
\setlength{\tabcolsep}{1.8pt}\footnotesize
\begin{tabular}{ccccccccccccc}
\hline
 &  & $ D_{1}^{CvM} $  & $ D_{2,1}^{CvM} $  & $ D_{2,2}^{CvM} $  & $ D_{1}^{KS} $  & $ D_{2,1}^{KS} $  & $D_{2,2}^{KS} $  & $ LBQ_1 $  & $ LBQ_2 $  & $ LBQ_3 $  & $ LBQ_{20} $  & $ LBQ_{25} $  \tabularnewline
\hline\hline

B1 & $10\%$  & $ 5.6 $  & $ 8.2 $  & $ 8.6 $  & $ 7.6 $  & $ 8.6 $  & $ 8.6 $  & $ 7.2 $  & $ 10.0 $  & $ 7.0 $  & $ 7.8 $  & $ 9.6 $  \tabularnewline
  & $5\%$    & $ 1.2 $  & $ 2.0 $  & $ 3.8 $  & $ 2.0 $  & $ 4.2 $  & $ 4.0 $  & $ 2.8 $  & $ 4.0 $  & $ 2.8 $  & $ 4.0 $  & $ 4.2 $  \tabularnewline
  & $1\%$    & $ 0.4 $  & $ 0.0 $  & $ 1.2 $  & $ 0.2 $  & $ 0.2 $  & $ 0.2 $  & $ 0.8 $  & $ 1.2 $  & $ 1.0 $  & $ 0.6 $  & $ 0.4 $  \tabularnewline
  \hline
B2& $10\%$   & $ 8.2 $  & $ 7.8 $  & $ 7.4 $  & $ 8.0 $  & $ 9.4 $  & $ 6.4 $  & $ 6.6 $  & $ 7.8 $  & $ 10.8 $  & $ 12.0 $  & $ 12.6 $  \tabularnewline
   & $5\%$   & $ 3.4 $  & $ 3.2 $  & $ 3.2 $  & $ 3.4 $  & $ 4.0 $  & $ 3.6 $  & $ 3.4 $  & $ 4.6 $  & $ 6.2 $  & $ 8.2 $  & $ 5.8 $  \tabularnewline
   & $1\%$   & $ 0.2 $  & $ 0.4 $  & $ 0.2 $  & $ 0.4 $  & $ 0.4 $  & $ 1.4 $  & $ 1.2 $  & $ 1.4 $  & $ 0.8 $  & $ 1.0 $  & $ 1.8 $  \tabularnewline
  \hline
B3 & $10\%$  & $ 8.2 $  & $ 8.0 $  & $ 9.0 $  & $ 8.4 $  & $ 6.8 $  & $ 9.0 $  & $ 6.6 $  & $ 11.2 $  & $ 11.6 $  & $ 9.8 $  & $ 9.2 $  \tabularnewline
  & $5\%$    & $ 3.4 $  & $ 2.2 $  & $ 3.8 $  & $ 3.0 $  & $ 2.4 $  & $ 3.6 $  & $ 3.6 $  & $ 6.0 $  & $ 6.0 $  & $ 6.0 $  & $ 5.0 $  \tabularnewline
  & $1\%$    & $ 0.8 $  & $ 0.8 $  & $ 0.2 $  & $ 0.4 $  & $ 0.4 $  & $ 0.0 $  & $ 1.0 $  & $ 0.6 $  & $ 1.0 $  & $ 1.8 $  & $ 2.0 $  \tabularnewline
  \hline
B4 & $10\%$  & $ 73.2 $  & $ 70.4 $  & $ 59.0 $  & $ 62.8 $  & $ 63.8 $  & $ 55.8 $  & $ 18.0 $  & $ 13.2 $  & $ 12.6 $  & $ 9.4 $  & $ 9.0 $  \tabularnewline
  & $5\%$    & $ 63.0 $  & $ 61.0 $  & $ 49.2 $  & $ 52.4 $  & $ 49.6 $  & $ 43.4 $  & $ 8.4 $  & $ 6.6 $  & $ 6.4 $  & $ 6.4 $  & $ 5.4 $  \tabularnewline
  & $1\%$    & $ 43.8 $  & $ 37.4 $  & $ 30.0 $  & $ 33.0 $  & $ 26.2 $  & $ 23.8 $  & $ 2.4 $  & $ 0.4 $  & $ 2.0 $  & $ 1.0 $  & $ 1.6 $  \tabularnewline
  \hline
B5 & $10\%$  & $ 72.0 $  & $ 69.8 $  & $ 63.6 $  & $ 61.8 $  & $ 64.4 $  & $ 61.6 $  & $ 18.6 $  & $ 14.2 $  & $ 12.2 $  & $ 10.6 $  & $ 10.4 $  \tabularnewline
  & $5\%$    & $ 63.8 $  & $ 61.0 $  & $ 53.8 $  & $ 45.2 $  & $ 55.2 $  & $ 44.8 $  & $ 9.4 $  & $ 6.0 $  & $ 8.0 $  & $ 5.8 $  & $ 4.6 $  \tabularnewline
  & $1\%$    & $ 38.6 $  & $ 38.0 $  & $ 25.2 $  & $ 32.0 $  & $ 29.8 $  & $ 27.6 $  & $ 3.0 $  & $ 3.4 $  & $ 1.2 $  & $ 0.8 $  & $ 0.8 $  \tabularnewline
  \hline
B6 & $10\%$  & $ 70.2 $  & $ 68.6 $  & $ 65.8 $  & $ 59.4 $  & $ 64.4 $  & $ 55.4 $  & $ 15.4 $  & $ 15.4 $  & $ 15.2 $  & $ 12.2 $  & $ 11.2 $  \tabularnewline
  & $5\%$    & $ 60.4 $  & $ 53.4 $  & $ 52.2 $  & $ 46.8 $  & $ 50.4 $  & $ 50.4 $  & $ 11.4 $  & $ 8.8 $  & $ 8.8 $  & $ 5.2 $  & $ 7.6 $  \tabularnewline
  & $1\%$    & $ 45.6 $  & $ 36.2 $  & $ 24.4 $  & $ 25.0 $  & $ 29.6 $  & $ 39.8 $  & $ 5.8 $  & $ 2.6 $  & $ 1.2 $  & $ 2.2 $  & $ 1.0 $  \tabularnewline
  \hline
 \end{tabular}
 \label{tbl:T100}
\end{table}

\begin{table}[t]
\caption{Rejection rates of tests for the Normal independent in time
composite null hypothesis defined in (\ref{eq:H0B})
and DGP-B1 - DGP-B6. Both Cramer-von Mises
and Kolmogorov-Smirnov norms are used.
Tests based on the one-parameter process, the two-parameter process with the
$1$ and $2$ lags, as well as Ljung-Box (Box et al., 1994) test statistics with
$1$, $2$, $3$, $20$, and $25$ lags,
are considered. Sample sizes are $T=300$. }
\centering
\setlength{\tabcolsep}{1.8pt}\footnotesize
\begin{tabular}{ccccccccccccc}
\hline
 &  & $ D_{1}^{CvM} $  & $ D_{2,1}^{CvM} $  & $ D_{2,2}^{CvM} $  & $ D_{1}^{KS} $  & $ D_{2,1}^{KS} $  & $D_{2,2}^{KS} $  & $ LBQ_1 $  & $ LBQ_2 $  & $ LBQ_3 $  & $ LBQ_{20} $  & $ LBQ_{25} $  \tabularnewline
\hline
\hline
B1 & $10\%$   & $ 10.2 $  & $ 9.6 $  & $ 8.0 $  & $ 7.8 $  & $ 10.0 $  & $ 11.0 $  & $ 12.4 $  & $ 10.6 $  & $ 13.8 $  & $ 16.2 $  & $ 13.6 $  \tabularnewline
  & $5\%$     & $ 4.4 $  & $ 4.0 $  & $ 4.2 $  & $ 3.2 $  & $ 5.6 $  & $ 4.0 $  & $ 7.8 $  & $ 5.2 $  & $ 6.4 $  & $ 7.0 $  & $ 5.0 $  \tabularnewline
  & $1\%$     & $ 0.6 $  & $ 0.8 $  & $ 0.4 $  & $ 0.0 $  & $ 0.8 $  & $ 0.2 $  & $ 1.6 $  & $ 0.2 $  & $ 0.4 $  & $ 0.4 $  & $ 1.0 $  \tabularnewline
  \hline
B2& $10\%$           & $ 9.0 $    & $ 10.2 $  & $ 10.0 $  & $ 7.6 $  & $ 9.2 $  & $ 8.4 $  & $ 9.6 $  & $ 6.4 $  & $ 8.4 $  & $ 8.4 $  & $ 10.0 $  \tabularnewline
   & $5\%$           & $ 4.0 $    & $ 4.2 $  & $ 4.6 $  & $ 2.8 $  & $ 4.0 $  & $ 4.0 $  & $ 3.4 $  & $ 3.6 $  & $ 4.8 $  & $ 5.2 $  & $ 3.8 $  \tabularnewline
   & $1\%$             & $ 0.2 $  & $ 1.0 $  & $ 0.6 $  & $ 0.4 $  & $ 1.6 $  & $ 0.6 $  & $ 0.8 $  & $ 0.8 $  & $ 1.0 $  & $ 1.6 $  & $ 1.2 $  \tabularnewline
  \hline
B3 & $10\%$    & $ 9.4 $  & $ 9.2 $  & $ 11.0 $  & $ 11.2 $  & $ 10.6 $  & $ 11.2 $  & $ 7.4 $  & $ 10.2 $  & $ 11.0 $  & $ 10.2 $  & $ 9.8 $  \tabularnewline
  & $5\%$      & $ 6.0 $  & $ 8.0 $  & $ 7.0 $  & $ 5.0 $  & $ 6.4 $  & $ 7.0 $  & $ 3.6 $  & $ 4.8 $  & $ 4.8 $  & $ 5.4 $  & $ 7.6 $  \tabularnewline
  & $1\%$      & $ 1.8 $  & $ 2.0 $  & $ 0.6 $  & $ 0.8 $  & $ 1.0 $  & $ 1.0 $  & $ 1.0 $  & $ 1.6 $  & $ 0.6 $  & $ 0.4 $  & $ 1.6 $  \tabularnewline
  \hline
B4 & $10\%$    & $ 98.4 $  & $ 97.6 $  & $ 97.2 $  & $ 96.2 $  & $ 98.6 $  & $ 95.8 $  & $ 24.2 $  & $ 16.2 $  & $ 16.2 $  & $ 12.0 $  & $ 11.6 $  \tabularnewline
  & $5\%$      & $ 97.8 $  & $ 97.0 $  & $ 95.0 $  & $ 94.8 $  & $ 95.4 $  & $ 91.4 $  & $ 17.4 $  & $ 8.0 $  & $ 9.2 $  & $ 4.2 $  & $ 4.8 $  \tabularnewline
  & $1\%$      & $ 95.2 $  & $ 90.6 $  & $ 90.2 $  & $ 83.6 $  & $ 87.8 $  & $ 80.8 $  & $ 5.4 $  & $ 1.8 $  & $ 1.0 $  & $ 1.2 $  & $ 0.2 $  \tabularnewline
  \hline
B5 & $10\%$     & $ 99.6 $  & $ 99.2 $  & $ 97.6 $  & $ 97.2 $  & $ 97.4 $  & $ 95.4 $  & $ 18.2 $  & $ 13.8 $  & $ 12.0 $  & $ 11.8 $  & $ 10.2 $  \tabularnewline
  & $5\%$       & $ 99.0 $  & $ 98.0 $  & $ 95.6 $  & $ 95.0 $  & $ 93.6 $  & $ 91.6 $  & $ 12.0 $  & $ 5.8 $  & $ 6.8 $  & $ 5.8 $  & $ 4.6 $  \tabularnewline
  & $1\%$       & $ 94.4 $  & $ 89.0 $  & $ 86.4 $  & $ 72.4 $  & $ 79.8 $  & $ 78.0 $  & $ 7.2 $  & $ 1.8 $  & $ 0.6 $  & $ 0.4 $  & $ 0.4 $  \tabularnewline
  \hline
B6 & $10\%$      & $ 99.0 $  & $ 99.0 $  & $ 96.8 $  & $ 98.2 $  & $ 97.6 $  & $ 95.6 $  & $ 19.4 $  & $ 15.8 $  & $ 14.2 $  & $ 12.6 $  & $ 10.4 $  \tabularnewline
  & $5\%$        & $ 98.0 $  & $ 97.6 $  & $ 95.0 $  & $ 94.4 $  & $ 95.4 $  & $ 90.6 $  & $ 13.0 $  & $ 8.2 $  & $ 9.0 $  & $ 5.4 $  & $ 4.4 $  \tabularnewline
  & $1\%$        & $ 91.4 $  & $ 92.2 $  & $ 86.0 $  & $ 79.2 $  & $ 86.4 $  & $ 84.6 $  & $ 5.2 $  & $ 2.8 $  & $ 1.2 $  & $ 1.8 $  & $ 0.8 $  \tabularnewline
  \hline
\end{tabular}
\label{tbl:T300}
\end{table}

\subsection{Part C} We study the finite sample performance of the test
statistics to test the null hypothesis that the series is a bivariate nonlinear
in mean model, as considered in our empirical application in
Section~\ref{sect:emp}.  The results are presented for sample sizes $T=100$ in
Table~\ref{tbl:C100} and $T=300$ in Table~\ref{tbl:C300}.

\begin{table}[t]
\caption{Rejection rates of tests for
the null hypothesis defined in (\ref{eq:H0NAR})
and DGP-C1 - DGP-C6. Both Cramer-von Mises
and Kolmogorov-Smirnov norms are used.
Tests based on the one-parameter process, the 
two-parameter process with $1$ and $2$ lags, as well as Ljung-Box test
statistics with $1$, $2$, $3$, $20$, and $25$ lags and Jarque-Bera test
statistics, are considered. Sample sizes are $T=100$. }

\centering
\setlength{\tabcolsep}{1.8pt}\footnotesize
\begin{tabular}{cccccccccccccc}
\hline
 &  & $ D_{1}^{CvM} $  & $ D_{2,1}^{CvM} $  & $ D_{2,2}^{CvM} $  & $ D_{1}^{KS} $  & $ D_{2,1}^{KS} $  & $D_{2,2}^{KS} $  & $ LBQ_1 $  & $ LBQ_2 $  & $ LBQ_3 $  & $ LBQ_{20} $  & $ LBQ_{25} $ & $ JB$  \tabularnewline
\hline
\hline
 C1  & $ 10\% $  & $ 8.9 $  & $ 10.7 $  & $ 10.1 $  & $ 9.3 $  & $ 11.4 $  & $ 9.7 $  & $ 9.2 $  & $ 9.6 $  & $ 8.3 $  & $ 9.3 $  & $ 8.5 $ & $ 10.9 $ \tabularnewline
  & $ 5\% $  & $ 4.8 $  & $ 4.2 $  & $ 5.3 $  & $ 5.9 $  & $ 4.6 $  & $ 5.5 $  & $ 4.2 $  & $ 4.2 $  & $ 3.5 $  & $ 4.5 $  & $ 4.0 $ & $ 6.4 $  \tabularnewline
   & $ 1\% $  & $ 1.2 $  & $ 0.7 $  & $ 1.0 $  & $ 1.5 $  & $ 1.2 $  & $ 0.5 $  & $ 0.7 $  & $ 0.4 $  & $ 0.3 $  & $ 0.9 $  & $ 0.6 $ & $ 1.3 $ \tabularnewline
    \hline
C2  & $ 10\% $  & $ 37.3 $  & $ 36.2 $  & $ 32.5 $  & $ 29.2 $  & $ 32.2 $  & $ 28.5 $  & $ 9.9 $  & $ 11.0 $  & $ 11.6 $  & $ 9.6 $  & $ 8.1 $ & $ 64.7 $  \tabularnewline
 & $ 5\% $  & $ 27.7 $  & $ 24.8 $  & $ 22.7 $  & $ 20.3 $  & $ 21.9 $  & $ 19.3 $  & $ 4.6 $  & $ 5.6 $  & $ 5.1 $  & $ 4.9 $  & $ 4.5 $ & $ 56.6  $ \tabularnewline
  & $ 1\% $  & $ 15.2 $  & $ 9.9 $  & $ 10.2 $  & $ 6.7 $  & $ 9.0 $  & $ 10.0 $  & $ 1.6 $  & $ 1.5 $  & $ 0.6 $  & $ 0.4 $  & $ 0.7 $  & $ 32.2 $\tabularnewline
   \hline
 C3  & $ 10\% $  & $ 62.1 $  & $ 57.0 $  & $ 48.7 $  & $ 51.0 $  & $ 53.0 $  & $ 46.5 $  & $ 18.5 $  & $ 19.1 $  & $ 16.3 $  & $ 13.8 $  & $ 12.3 $ & $ 83.6 $  \tabularnewline
     & $ 5\% $  & $ 49.1 $  & $ 44.1 $  & $ 41.5 $  & $ 37.9 $  & $ 41.3 $  & $ 35.9 $  & $ 11.4 $  & $ 11.7 $  & $ 8.8 $  & $ 6.9 $  & $ 5.5 $ & $ 80.2  $  \tabularnewline
      & $ 1\% $  & $ 29.8 $  & $ 28.8 $  & $ 23.9 $  & $ 13.1 $  & $ 23.8 $  & $ 16.7 $  & $ 4.7 $  & $ 2.1 $  & $ 2.9 $  & $ 2.8 $  & $ 2.3 $ & $ 62.9  $ \tabularnewline
       \hline
  C4  & $ 10\% $  & $ 93.9 $  & $ 91.3 $  & $ 90.8 $  & $ 86.6 $  & $ 87.1 $  & $ 85.8 $  & $ 29.2 $  & $ 26.5 $  & $ 21.9 $  & $ 16.1 $  & $ 13.9 $ & $ 97.5 $ \tabularnewline
         & $ 5\% $  & $ 89.4 $  & $ 87.3 $  & $ 85.4 $  & $ 79.6 $  & $ 81.6 $  & $ 79.4 $  & $ 22.0 $  & $ 19.2 $  & $ 14.6 $  & $ 8.0 $  & $ 7.6 $ & $ 97.1 $ \tabularnewline
          & $ 1\% $  & $ 79.0 $  & $ 79.3 $  & $ 73.6 $  & $ 60.1 $  & $ 66.3 $  & $ 59.7 $  & $ 13.0 $  & $ 9.9 $  & $ 6.5 $  & $ 2.7 $  & $ 1.3 $ & $ 92.3 $ \tabularnewline
           \hline
 C5  & $ 10\% $  & $ 11.2 $  & $ 30.2 $  & $ 10.1 $  & $ 10.0 $  & $ 23.2 $  & $ 8.3 $  & $ 81.6 $  & $ 68.5 $  & $ 47.7 $  & $ 26.1 $  & $ 24.7 $  & $ 11.6 $  \tabularnewline
  & $ 5\% $  & $ 4.4 $  & $ 19.3 $  & $ 5.8 $  & $ 4.6 $  & $ 13.2 $  & $ 4.1 $  & $ 71.0 $  & $ 52.9 $  & $ 34.0 $  & $ 17.7 $  & $ 15.0 $  & $ 4.6 $  \tabularnewline
   & $ 1\% $  & $ 1.1 $  & $ 7.6 $  & $ 1.0 $  & $ 1.2 $  & $ 4.3 $  & $ 0.9 $  & $ 52.3 $  & $ 23.6 $  & $ 9.5 $  & $ 5.8 $  & $ 3.8 $  & $ 1.0 $  \tabularnewline
    \hline
     C6  & $ 10\% $  & $ 10.4 $  & $ 60.4 $  & $ 8.6 $  & $ 10.4 $  & $ 38.7 $  & $ 9.2 $  & $ 99.9 $  & $ 97.2 $  & $ 93.2 $  & $ 61.8 $  & $ 54.1 $  & $ 10.8 $  \tabularnewline
      & $ 5\% $  & $ 6.3 $  & $ 44.0 $  & $ 4.5 $  & $ 5.3 $  & $ 26.3 $  & $ 5.3 $  & $ 99.2 $  & $ 94.9 $  & $ 84.8 $  & $ 48.8 $  & $ 41.8 $  & $ 5.0 $  \tabularnewline
       & $ 1\% $  & $ 0.7 $  & $ 22.7 $  & $ 0.7 $  & $ 1.2 $  & $ 6.5 $  & $ 0.7 $  & $ 96.0 $  & $ 82.1 $  & $ 56.4 $  & $ 21.7 $  & $ 13.1 $  & $ 1.2 $  \tabularnewline
        \hline
\end{tabular}
\label{tbl:C100}
\end{table}

\begin{table}[t]
\caption{Rejection rates of tests for 
the null hypothesis defined in (\ref{eq:H0NAR})
and DGP-C1 - DGP-C6. Both Cramer-von Mises
and Kolmogorov-Smirnov norms are used.
Tests based on the one-parameter process,
the two-parameter process with $1$ and $2$ lags, 
as well as Ljung-Box test statistics with $1$, $2$, $3$, $20$, and $25$ lags
and Jarque-Bera test statistics, are considered. Sample sizes are $T=300$. }
\centering
\setlength{\tabcolsep}{1.8pt}\footnotesize
\begin{tabular}{cccccccccccccc}
\hline
 &  & $ D_{1}^{CvM} $  & $ D_{2,1}^{CvM} $  & $ D_{2,2}^{CvM} $  & $ D_{1}^{KS} $  & $ D_{2,1}^{KS} $  & $D_{2,2}^{KS} $  & $ LBQ_1 $  & $ LBQ_2 $  & $ LBQ_3 $  & $ LBQ_{20} $  & $ LBQ_{25} $ & $ JB $ \tabularnewline
\hline\hline
 C1  & $ 10\% $  & $ 8.3 $  & $ 10.4 $  & $ 8.7 $  & $ 8.3 $  & $ 9.0 $  & $ 7.8 $  & $ 11.5 $  & $ 9.8 $  & $ 8.8 $  & $ 12.3 $  & $ 11.6 $ & $ 8.3 $ \tabularnewline
  & $ 5\% $  & $ 5.2 $  & $ 6.3 $  & $ 4.3 $  & $ 4.8 $  & $ 4.1 $  & $ 4.0 $  & $ 6.7 $  & $ 4.0 $  & $ 4.9 $  & $ 4.8 $  & $ 6.4 $ & $ 3.6 $ \tabularnewline
   & $ 1\% $  & $ 0.6 $  & $ 0.8 $  & $ 0.7 $  & $ 0.9 $  & $ 0.8 $  & $ 0.5 $  & $ 1.0 $  & $ 1.2 $  & $ 1.1 $  & $ 1.3 $  & $ 1.9 $  & $ 0.8 $\tabularnewline
    \hline
    C2  & $ 10\% $  & $ 85.1 $  & $ 83.1 $  & $ 76.5 $  & $ 74.2 $  & $ 74.7 $  & $ 70.3 $  & $ 10.5 $  & $ 10.2 $  & $ 10.5 $  & $ 13.2 $  & $ 11.3 $ & $ 96.6 $  \tabularnewline
     & $ 5\% $  & $ 77.4 $  & $ 73.2 $  & $ 66.4 $  & $ 61.7 $  & $ 63.5 $  & $ 63.4 $  & $ 6.3 $  & $ 5.2 $  & $ 5.5 $  & $ 7.3 $  & $ 6.6 $  & $ 94.7 $ \tabularnewline
      & $ 1\% $  & $ 58.5 $  & $ 43.3 $  & $ 43.5 $  & $ 35.8 $  & $ 37.7 $  & $ 32.1 $  & $ 2.6 $  & $ 1.0 $  & $ 1.9 $  & $ 1.7 $  & $ 0.9 $ & $ 88.7 $ \tabularnewline
       \hline
       C3  & $ 10\% $  & $ 97.3 $  & $ 97.0 $  & $ 96.0 $  & $ 93.0 $  & $ 94.5 $  & $ 93.0 $  & $ 23.0 $  & $ 21.5 $  & $ 15.4 $  & $ 11.8 $  & $ 13.0 $ & $ 99.5 $ \tabularnewline
        & $ 5\% $  & $ 95.6 $  & $ 95.1 $  & $ 93.2 $  & $ 87.4 $  & $ 91.0 $  & $ 85.5 $  & $ 15.6 $  & $ 11.5 $  & $ 9.2 $  & $ 6.9 $  & $ 7.7 $ & $ 99.3 $ \tabularnewline
         & $ 1\% $  & $ 89.6 $  & $ 90.0 $  & $ 83.4 $  & $ 69.1 $  & $ 75.6 $  & $ 64.6 $  & $ 6.0 $  & $ 3.3 $  & $ 2.4 $  & $ 3.2 $  & $ 1.7 $ & $ 98.3 $ \tabularnewline
          \hline
           C4  & $ 10\% $  & $ 100.0 $  & $ 100.0 $  & $ 100.0 $  & $ 100.0 $  & $ 100.0 $  & $ 100.0 $  & $ 40.7 $  & $ 36.1 $  & $ 31.2 $  & $ 17.9 $  & $ 17.8 $ & $ 100$  \tabularnewline
            & $ 5\% $  & $ 100.0 $  & $ 100.0 $  & $ 100.0 $  & $ 100.0 $  & $ 100.0 $  & $ 100.0 $  & $ 34.2 $  & $ 28.8 $  & $ 22.4 $  & $ 12.8 $  & $ 10.0 $ & $ 100 $ \tabularnewline
             & $ 1\% $  & $ 100.0 $  & $ 100.0 $  & $ 100.0 $  & $ 99.8 $  & $ 100.0 $  & $ 99.7 $  & $ 22.6 $  & $ 18.1 $  & $ 10.3 $  & $ 3.9 $  & $ 3.4 $  & $ 100 $ \tabularnewline
              \hline
 C5  & $ 10\% $  & $ 11.1 $  & $ 73.6 $  & $ 11.2 $  & $ 10.0 $  & $ 44.6 $  & $ 10.7 $  & $ 99.8 $  & $ 99.0 $  & $ 97.4 $  & $ 77.9 $  & $ 73.9 $  & $ 9.2 $  \tabularnewline
  & $ 5\% $  & $ 6.4 $  & $ 61.9 $  & $ 6.0 $  & $ 5.7 $  & $ 31.3 $  & $ 4.8 $  & $ 99.4 $  & $ 97.8 $  & $ 94.8 $  & $ 66.3 $  & $ 59.0 $  & $ 4.2 $  \tabularnewline
   & $ 1\% $  & $ 1.6 $  & $ 33.8 $  & $ 0.7 $  & $ 1.7 $  & $ 16.0 $  & $ 0.9 $  & $ 97.6 $  & $ 88.7 $  & $ 81.6 $  & $ 40.9 $  & $ 37.1 $  & $ 0.7 $  \tabularnewline
    \hline
     C6  & $ 10\% $  & $ 9.1 $  & $ 99.5 $  & $ 9.5 $  & $ 9.0 $  & $ 84.4 $  & $ 8.5 $  & $ 100.0 $  & $ 100.0 $  & $ 100.0 $  & $ 99.8 $  & $ 99.7 $  & $ 8.7 $  \tabularnewline
      & $ 5\% $  & $ 5.1 $  & $ 98.6 $  & $ 4.8 $  & $ 5.3 $  & $ 71.3 $  & $ 4.3 $  & $ 100.0 $  & $ 100.0 $  & $ 100.0 $  & $ 99.5 $  & $ 99.3 $  & $ 3.9 $  \tabularnewline
       & $ 1\% $  & $ 2.0 $  & $ 90.6 $  & $ 0.9 $  & $ 2.1 $  & $ 49.5 $  & $ 0.3 $  & $ 100.0 $  & $ 100.0 $  & $ 100.0 $  & $ 98.7 $  & $ 97.4 $  & $ 0.5 $  \tabularnewline
       \hline
\end{tabular}
\label{tbl:C300}
\end{table}

In DGP-C1, $Y_t$ is generated by the model considered in
Section~\ref{sect:emp}, with $b$ and $\Sigma$  equal to the estimate obtained
therein.  Rejection rates deliver the empirical size of the tests, which is
close to nominal for both sample sizes. 

The power of the tests is studied against Student-$t$ models with the same
dynamics.  The closest to normal, $t_7$ model is rejected between $19.3\%$ for
KS-test with lag $2$ and $27.7\%$ for CvM-test with no lags for sample size
$T=100$.  The rejected rates are doubled for $t_5$ and $4$ times larger for
$t_3$. For $T=300$, rejection rates are between $61.7\%$ and $77.4\%$ for
$t_7$, around $90\%$ for $t_5$, and $100\%$ for $t_3$. Note that linear
correlation tests are unlikely to reject any of these alternatives. If one knew
that the misspecification comes solely from the skewed marginals, then a
moment-based test would be enough. Indeed, the Jarque-Bera test rejects $97\%$
even for $t_7$ with $T=300$.    

We also consider power in case of the dynamic misspecification.  Let $Y_t$ be
generated by the model considered in Section~\ref{sect:emp}, with $b$ and
$\Sigma$  equal to the estimates obtained therein, except that the conditional
mean for the output has an additional lag $\tilde\mu_{t1}=\mu_{t1}+ \alpha
Y_{t-2,2}$, where $\alpha=0.5$ for DGP-C5 and $\alpha=0.9$ for DGP-C6. For
$T=100$, the linear correlation tests behave very well, while the empirical
process-based tests require more data: the rejection rate is only $44\%$ for
the CvM-test based on the $1$-lag two-parameter process. The rejection rates
become  $98.6\%$ for $T=300$ for CVM and $71.3\%$ for the KS-test. Note, that
the one-parameter empirical process-based tests do not have power against
dynamic misspecification.

Based on the simulation results in this section, we suggest to practitioners
that they first perform standard tests (e.g., correlation and moment-based). If
the model is rejected, then the source of the misspecification is identified as
usual.  If the model is not rejected, it could be that the standard tests do
not have power. Then they should complement the analysis with the tests
proposed in this paper.

\section{Conclusion}

In this paper we discuss how to test a multivariate conditional distribution
against a wide set of alternatives. Our method is a formalization of the
approach for density forecast evaluation proposed by Diebold et al. (1999), and
is based on certain empirical processes of random variables obtained by
applying the dynamic Rosenblatt transform. We discuss the importance of
cross-sectional and dynamic checks and compare our strategy to other commonly
used nonparametric tests. We state the asymptotic properties of the tests under
the null and local alternatives and present a bootstrap justification. We
study the size and power properties of the test in finite samples for basic
bilinear dynamic models. For $T=100$ the tests are slightly undersized, but for
$T=300$ the situation improves. From power experiments, we can confirm the
importance of using tests based in multi-parameter empirical processes if we
want to cover different alternatives. Finally, the tests are applied to the
real UK macroeconomic data.

\section{Acknowledgments}

I am grateful to Pentti Saikkonen, Timo Terasvirta and Carlos Velasco for
important suggestions, and to Miguel Delgado, Manuel Dominguez, Jesus Gonzalo,
Ignacio Lobato, Juan Mora, Stefan Sperlich, Abderrahim Taamouti and the
anonymous referees for helpful comments.  I thank Oxana Budjko, Patrick Kelly,
Dmitry Makarov and Anton Suvorov for their hospitality during my visits to the
New Economic School and Higher School of Economics in Moscow.  Financial
support from the Spanish Ministerio de Economia y Competitividad (grants
ECO2017-86009-P and ECO2014-57007p) is gratefully acknowledged.

\appendix{} \section{Appendix}
\subsection{Asymptotic properties} \label{sect:asy} In this section we
formulate the asymptotic properties of the proposed test statistics. We start
with the simple case of when we know parameters, then we study how  the
asymptotic distribution changes if we estimate parameters. We provide the
analysis under the null and under the local and fixed alternatives. We will
need assumptions on conditional dfs, the form of the parametric family of dfs,
and on the estimator.  The theory relies on the weak convergence result for
multivariate empirical processes proved in Kheifets (2015), which notation we
use. Extension from a univariate case to a multivariate case is obtained by
noting that under the null we obtain, in both cases, a series of univariate
i.i.d.~standard uniform random variables and results on the parameter
estimation effect are due to smoothness of the conditional distribution with
 respect to the parameter. Therefore we omit the proofs.

\noindent {\textbf{Assumption 1.} The conditional distributions 
$F_{t}(y|\Omega _{t},\theta )$ are absolutely continuous.}

The following proposition provides the main result about the PIT. The idea goes
back to Rosenblatt (1952).

\begin{prop}
\label{propFY} Suppose Assumption 1 holds. Then, under $H_0$, the random
variables $\{U_k\}_{k=1}^n$ are i.i.d.\ standard uniform.
\end{prop}

We first describe the asymptotic behavior of the process $V_{2T}$ under
$H_{0}$. Denote by ``$\implies $''weak convergence of stochastic processes as
random elements of the Skorokhod space $D\left( [0,1]^{2}\right) $. Since
dimension $d$ is fixed, all the asymptotic results $T\rightarrow \infty $ can
be formulated in terms of $n\rightarrow \infty .$

\begin{prop}
\label{proplimitV} Suppose Assumption 1 holds. Then, under $H_{0}$,
\begin{equation*}
V_{2T}\implies V_{2\infty },
\end{equation*}
where $V_{2\infty }({r})$ is bi-parameter zero-mean Gaussian process with
covariance
\begin{equation}
\mathop{\rm Cov}\nolimits_{V_{2\infty }}({r},{s})=(r_{1}\wedge
s_{1})(r_{2}\wedge s_{2})+(r_{1}\wedge s_{2})r_{2}s_{1}+(r_{2}\wedge
s_{1})r_{1}s_{2}-3r_{1}r_{2}s_{1}s_{2}.  \label{eqasscov}
\end{equation}
\end{prop}

By applying the Continuous Mapping Theorem, we can obtain the asymptotic
distribution of $D_{2T}$ and other statistics, with critical values that can
be simulated by Monte Carlo and tabulated.

In case of a composite null hypothesis, estimating the parameter may affect the
asymptotic distribution under the null, see  Durbin (1973). We will take this
into
account, i.e., we derive the difference between $\hat{V}_{2T}({r})$ and 
$V_{2T}({r})$. 
 Let $\Vert \cdot \Vert $ denote
the Euclidean norm for matrices, i.e., $\Vert A\Vert =\sqrt{\mathop{\rm tr}
\nolimits(A^{\prime }A)}$, and for $\varepsilon >0,$ $B(a,\varepsilon )$\ is
an open ball in $R^{L}$ with the center at point $a$ and radius 
$\varepsilon $. In particular, for some $M>0$, denote $B_{T}=B\left( \theta
_{0},M T^{-1/2}\right) =\{\theta :||\theta -\theta _{0}||\leq M T^{-1/2}\}$.
Let $\eta_k\left(r,u,v\right)=F_{Z_k}\left( F_{Z_k}^{-1}\left(r |u\right)
|v\right)$. The following assumption from Kheifets (2015) ensures 
smoothness of the distribution and is stated in terms of $Z_k$. 

\asss{2}{}
\begin{itemize}
\item[(2.1)] 
\begin{equation*}
E\sup_{t=1,..,n}\sup_{u\in B_n}\sup_{r\in [0,1]}
\left\vert \eta _{t}\left(r,u,\theta_0\right)-r \right\vert 
=O\left( n^{-1/2}\right).
\end{equation*}
\item[(2.2)]  
$\forall M\in(0,\infty)$, $\forall M_2\in(0,\infty)$, and $\forall\delta> 0$,
\begin{equation*}
\sup_{r\in [0,1]}
\frac{1}{\sqrt{n}}
\sum_{t=1}^{n}
\sup_{||u-v||\le M_2 n^{-1/2-\delta}}
\left|
{\eta
}_{t}\left( r,u,\theta_0\right)
-
{\eta }_{t}\left(
r,v,\theta_0\right)
\right| 
=o_{p}\left(1\right).
\end{equation*}
\item[(2.3)]  
$\forall M\in(0,\infty)$, $\forall M_2\in(0,\infty)$, and $\forall\delta> 0$,
\begin{equation*}
\sup_{|r-s|\le M_2 n^{-1/2-\delta} }
\frac{1}{\sqrt{n}}
\sum_{t=1}^{n}
\sup_{u\in B_n}
\left|
{\eta
}_{t}\left(r, u,\theta_0\right)
-
{\eta }_{t}\left(s, u,\theta_0\right)
\right| 
=o_{p}\left(1\right).
\end{equation*}
\item[(2.4)]  $\forall M\in(0,\infty)$, 
there exists a uniformly continuous (vector) function $h(r)$
from $[0,1]^{2}$ to $R^{L}$ such that
\begin{equation*}
\sup_{u,v\in B_n }\sup_{r\in \lbrack
0,1]^{2}}\left\vert
\frac{1}{\sqrt{n}}\sum_{t=2}^{n}h_t-h(r)^{\prime }{\sqrt{n}\left( u-v\right) }
\right\vert =o_{p}(1),
\end{equation*}
where
\begin{equation*}
h_t=
\left(\eta_{t-1}\left( r_{2},u,v\right)
-r_{2}\right) r_{1} 
+\left( \eta_{t}\left( r_{1},u,v\right) -r_{1}\right)
I\left( U_{t-1}\leq \eta_{t-1}\left(  r_{2},u,v\right)\right).
\end{equation*}
\end{itemize}

If the cdf $F_{Z_{k}}\left( x|\theta \right) $ is continuously differentiable
with respect to $\theta$ (uniformly in $k$ and $x$), with bounded derivatives,
then the mean value theorem will guarantee Assumption 2 under some additional
regularity conditions; see the discussion in Kheifets (2015).

We need to assume the existence of a linear expansion of the estimator.

\noindent {\textbf{Assumption 3.} When the sample is generated by the null 
$F_{t}(y|\Omega _{t},\theta _{0})$, the estimator $\hat{\theta}$ admits a
linear expansion
\begin{equation}
\sqrt{T}(\hat{\theta}-\theta _{0})=\frac{1}{\sqrt{T}}\sum_{t=1}^{T}\psi
(\Omega _{t})l(U_{t})+o_{p}(1),  \label{eqestlinear}
\end{equation}
with $E_{F_{t}}\left( l(U_{t})|\Omega _{t}\right) =0$ and
\begin{equation*}
\frac{1}{T}\sum_{t=1}^{T}\psi (\Omega _{t})\psi (\Omega _{t})^{\prime
}l^{2}(U_{t})\overset{p_{F_{t}}}{\rightarrow }\Psi .
\end{equation*}
}

This assumption is satisfied for ML and nonlinear least square (NLS) estimators
under minor additional conditions. It will allow application of the CLT for
vector r.v. 
$({V_{2T}(r),}\frac{1}{\sqrt{T}}\sum_{t=1}^{T}\psi (\Omega
_{t})l(U_{t}))^{\prime }$.
Define
\begin{equation*}
C_{T}(r,s,\theta )=E\left(
\begin{array}{c}
{V_{2T}(r)} \\
\frac{1}{\sqrt{T}}\sum_{t=1}^{T}\psi (\Omega _{t})l(U_{t})
\end{array}
\right) \left(
\begin{array}{c}
{V_{2T}(s)} \\
\frac{1}{\sqrt{T}}\sum_{t=1}^{T}\psi (\Omega _{t})l(U_{t})
\end{array}
\right) ^{\prime }
\end{equation*}
and let $(V_{2\infty }(r),\psi _{\infty }^{\prime })^{\prime }$ be a zero-mean
Gaussian process with covariance function 
$C(r,s,\theta_{0})=\lim_{T\rightarrow \infty }C_{T}(r,s,\theta _{0})$.
Dependence on $\theta$ on the right-hand side (rhs) comes through $U_{t}$,
since they are obtained with PIT. Let ``$\overset{d}{\rightarrow }$'' denote
convergence in distribution. The following proposition establishes the limiting
distribution of our test statistics.

\begin{prop}
\label{propnull}Suppose Assumptions 1-3 hold. Then, under $H_{0}$,
\begin{equation*}
\Gamma (\hat{V}_{2T})\overset{d}{\rightarrow }\Gamma (\hat{V}_{2\infty }),
\end{equation*}
where
\begin{equation*}
\hat{V}_{2\infty }({r})=V_{2\infty }(r)-h(r)^{\prime }\psi _{\infty }.
\end{equation*}
\end{prop}

We now study the asymptotics under the sequence of local alternatives.
Suppose the conditional distribution function $H_{t}(y|\Omega _{t})$ is not
in the parametric family $F_{t}(y|\Omega _{t},\theta )$, i.e., for each $\theta
\in \Theta $ there exists $y\in R$ and one $\Omega _{t_{0}}$ that occurs with
positive probability, and
$H_{t_{0}}(y|\Omega _{t_{0}})\neq F_{t_{0}}(y|\Omega _{t_{0}},\theta )$. For
any $0<\delta /\sqrt{T}<1$, define the conditional cdf
\begin{equation*}
G_{Tt}(y|\Omega _{t},\theta )=\left( 1-\frac{\delta }{\sqrt{T}}\right)
F_{t}(y|\Omega _{t},\theta )+\frac{\delta }{\sqrt{T}}H_{t}(y|\Omega _{t}).
\end{equation*}
Now we define the local alternatives. \bigskip

$H_{1T}$: The conditional df of $Y_{t}$ is equal to $G_{Tt}(y|\Omega _{t},\theta
_{0})$ .

\bigskip To derive the asymptotic distribution of our test statistics under
the sequence of local alternatives, we need an assumption on $H_{t}(y|\Omega
_{t})$ similar to Assumption 1.

\noindent{\textbf{Assumption 4.} The conditional dfs $H_{t}(y|\Omega _{t})$
are continuous and strictly increasing in $y$.}

Under Assumptions 1 and 2, we also have that the conditional dfs 
$G_{Tt}(y|\Omega _{t},\theta )$ are continuously differentiable with respect
to~$\theta $, and continuous and strictly increasing in $y$. Under the
alternative, we will require that the estimator converges in probability.

\noindent{\textbf{Assumption 5.} $\hat{\theta}\overset{p}{\rightarrow }
\theta _{1}$ for some $\theta _{1}\in \Theta .$}

Under the null, together with Assumption 3, this would imply $\theta _{1}=
\theta _{0}$. Otherwise this is not necessary true, and $\theta _{1}$ is
often referred a ``pseudo-true'' value. In the next proposition we provide the
asymptotic distribution of our statistics under local and fixed alternatives. 
\begin{prop}
\label{propalt} Suppose Assumptions 1-5 hold. Then, under $H_{1T}$,
\begin{equation*}
\Gamma (\hat{V}_{2T}(r))\overset{d}{\rightarrow }\Gamma (\hat{V}_{2\infty
}(r)+\delta k(r)-\delta h(r)^{\prime }\mu ),
\end{equation*}%
where the drift is 
\begin{eqnarray*}
k(r) &=&\limfunc{plim}\frac{1}{n}\sum_{k=2}^{n}\left\{ \left[
H_{Z_{k-1}}(F_{Z_{k-1}}^{-1}(r_{2}|\Omega^{Z} _{k-1},\theta _{0})|\Omega^{Z}
_{k-1})-r_{2} \right] r_{1}\right. \\
&&\left. +\left[ H_{Z_{k}}(F_{Z_{k}}^{-1}(r_{1}|\Omega^{Z} _{k},\theta
_{0})|\Omega^{Z} _{k})-r_{1}\right] I\left( U_{k-1}\leq
H_{Z_{k-1}}(F_{Z_{k-1}}^{-1}(r_{2}|\Omega^{Z} _{k-1},\theta _{0})|\Omega^{Z}
_{k-1})\right) \right\},
\end{eqnarray*}
and
\begin{equation*}
\mu =\limfunc{plim}\frac{1}{n}\sum_{k=1}^{n}\psi (\Omega _{k})l(U_{k}).
\end{equation*}
\end{prop}
The first part of the drift is nonzero only if marginals do not coincide, while
the second part checks dependence structure.

\noindent{\textbf{Assumption 6.}
For all nonrandom sequences $\{\theta _{T}:T\geq 1\}$ for
which $\theta _{T}\rightarrow \theta _{0}$, we have
\begin{equation*}\label{eq:linexptr}
\sqrt{T}(\hat{\theta}-\theta_{T})=
\frac{1}{\sqrt{T}}\sum_{t=1}^{T}\psi\left(Y_{Tt},\Omega_{Tt},\theta_T\right)
  +o_{p}(1),
\end{equation*}
under $\{\theta _{T}:T\geq 1\}$, where 
\begin{equation*}
\frac{1}{T}\sum_{t=1}^{T}
\psi\left(Y_{Tt},\Omega_{Tt},\theta_T\right)
\psi\left(Y_{Tt},\Omega_{Tt},\theta_T\right)^{\prime}
\stackrel{p}{\rightarrow }\Psi .
\end{equation*}

The next proposition states that the bootstrap is first-order asymptotically
valid, since the asymptotic distribution of the test statistics under $\{\theta
_{T}:T\geq 1\} $  coincides with that of under the null, see Corollary 1 in
Andrews (1997).

\begin{prop}
\label{propboot}Suppose Assumptions 1-6 hold. Then, under $H_{1T}$, for any
nonrandom sequence $ \{\theta _{T}:T\geq 1\}$ for which $\theta _{T}\rightarrow
\theta _{0}$,
under $\{\theta _{T}:T\geq 1\}$,
\begin{equation*}\label{eq:GVB}
\Gamma (\hat{V}_{2T}({r}))\stackrel{d}{\rightarrow }\Gamma (\hat{V}_{2\infty
}({r})).
\end{equation*}
\end{prop}

\subsection{Rosenblatt transforms}\label{sect:transform}

We now show how to obtain the explicit formulas for the Rosenblatt transform.
Consider bivariate independent across-time series that are not necessary
identically distributed, i.e., $Y_{t}\in R^{2}$ with 
mean $\mu _{t}=\left( \mu_{t1},\mu_{t2}\right) ^{\prime }$ and covariance matrix
\begin{equation*}\Sigma_t =\left(
\begin{array}{cc}
\Sigma _{t11} & \Sigma _{t12} \\ 
\Sigma _{t12} & \Sigma _{t22}
\end{array}
\right).
\end{equation*}
Project $Y_{t2}$ on $Y_{t1}$, $Y_{t2}$ $=\beta
Y_{t1}+Y_{t1}^{\perp }$. The least square estimator gives $\beta =\frac{\Sigma
_{t12}}{\Sigma _{t11}}$, $Y_{t1}\perp Y_{t1}^{\perp }$, $E\left(
Y_{t1}^{\perp }\right) =\mu _{t2}-\frac{\Sigma _{t12}}{\Sigma _{t11}}\mu
_{t1},\ $and $Var\left( Y_{t1}^{\perp }\right) =\Sigma _{t22}-\frac{\Sigma
_{t22}^{2}}{\Sigma _{t11}}$. Under normality, one-dimensional unconditional
and conditional distributions are also normal, orthogonality coincides with
independence, and the distribution of the projection $Y_{t1}^{\perp }$ gives
us the desired conditional distribution $Y_{t2}$ on $Y_{t1}$ and all the past $
\Omega _{t}$. Then
\begin{eqnarray*}
U_{2\left( t-1\right) +1} &=&\Phi \left( \frac{Y_{t1}-\mu _{t1}}{\sqrt{
\Sigma _{t11}}}\right) \\
\ U_{2\left( t-1\right) +2} &=&\Phi \left( \frac{Y_{t1}^{\perp }-E\left(
Y_{t1}^{\perp }\right) }{\sqrt{Var\left( Y_{t1}^{\perp }\right) }}\right)
=\Phi \left( \frac{\left( Y_{t2}-\frac{\Sigma _{t12}}{\Sigma _{t11}}
Y_{t1}\right) -\left( \mu _{t2}-\frac{\Sigma _{t12}}{\Sigma _{t11}}\mu
_{t1}\right) }{\sqrt{\Sigma _{t22}-\frac{\Sigma _{t22}^{2}}{\Sigma _{t11}}}}
\right)
\end{eqnarray*}
are uniform i.i.d.~under the null, where $\Phi \left( \cdot \right) $ is
normal cdf. These formulas also simplify the likelihood calculation and can be
used for estimation purposes (see Tsay 2010, Section 10.3.2, and Bai and Chen
2008, Equation (16)). Another multivariate distribution often used to model
macro/finance data is a multivariate $t$-distribution. Since conditional
and subvector distributions belong to the same class for the multivariate
$t$-distribution, formulas for the Rosenblatt transform can be derived in
similar way, see Bai and Chen (2008, Equation (20)) for details.

\subsection{Verifying assumptions}\label{sect:verifya} 

Explicit formulas for the Rosenblatt transform derived in \ref{sect:transform}
show that the conditional distribution of $Z_k$,
$F_{Z_k}(z|\Omega_k^{Z},\theta)$ is normal with its mean and variance being
simple transformations of the conditional variables and parameters. Then, if
the model is smooth with respect to the argument and parameter, so is the
conditional distribution of $Z_k$; in particular, for the multivariate normal
models considered in the paper,
$\sup_{\theta,z,k}||\nabla_{\theta}F_{Z_k}(z|\Omega_k^{Z},\theta)||=O_p(1)$ and
$\sup_{\theta,z,k}f_{Z_k}(z|\Omega_k^{Z},\theta)=O_p(1)$. The first bound is
sufficient for Assumptions (2.1) and (2.2), while the second bound is
sufficient for Assumptions 1 and (2.3). Assumption (2.4) follows from the
uniform law of large numbers for stationary and ergodic time series since the
models are continuous in the parameter. Assumptions~3,~5, and~6 are standard
and are satisfied for MLE (see Andrews 2002). Finally, for alternatives
considered in Section~\ref{sect:mc}, either for dynamic or marginal
misspecification, Assumption 4 holds.  

Establishing conditions under which there exists a stationary and ergodic
solution to the bivariate LSTAR system considered here is beyond the scope of
this paper; however, the results of Saikkonen (2008) may be useful to achieve
this goal. 

\subsection{Computational details}\label{sect:comp} 

It takes 15 minutes to calculate bootstrapped critical values for the
nonsimulated bivariate series of length $T=159$ with $B=1000$ bootstrap
repetitions for the LSTAR model considered in Section 3. Test statistics
proposed in the paper are calculated in C. All other calculations are
implemented in R version 3.1.1, R Core Team (2017). In particular, functions
mclapply() from the \textbf{parallel} package and optim() from the
\textbf{stats} package were used. The calculations are made on a 3.6 GHz Intel
Core i7-4790 under Debian 3.16.


\renewcommand{\baselinestretch}{0.8}

\end{document}